\documentclass{jfm}
\usepackage{amsmath,stmaryrd,natbib,upgreek,graphicx}

\begin{document}

\newtheorem{lemma}{Lemma}
\newtheorem{corollary}{Corollary}

\shorttitle{Nonlinear small-deformation theory for droplet electrohydrodynamics} 
\shortauthor{D. Das and D. Saintillan} 

\title{A nonlinear small-deformation theory for  transient droplet electrohydrodynamics}

\author
 {
 Debasish Das
\and
 David Saintillan
 \corresp{\email{dstn@ucsd.edu}}
 }

\affiliation
{
Department of Mechanical and Aerospace Engineering, University of California San Diego, \\ 9500 Gilman Drive, La Jolla, CA 92093, USA
}

\maketitle

\begin{abstract}
The deformation of a viscous liquid droplet suspended in another liquid and subject to an applied electric field is a classic multiphase flow problem  best described by the Melcher-Taylor leaky dielectric model. The main assumption of the model is that any net charge in the system is concentrated on the interface between the two liquids as a result of the jump in Ohmic currents from the bulk. Upon application of the field, the drop can either attain a steady prolate or oblate shape with toroidal circulating flows both inside and outside arising from tangential stresses on the interface due to action of the field on the surface charge distribution. Since the pioneering work of \cite{taylor1966}, there have been numerous computational and theoretical studies to predict the deformations measured in experiments. Most existing theoretical models, however, have either neglected transient charge relaxation or nonlinear charge convection by the interfacial flow. In this work, we develop a novel small-deformation theory accurate to second order in electric capillary number ${O}(Ca_E^2)$ for the complete Melcher-Taylor model that includes transient charge relaxation, charge convection by the flow, as well as transient shape deformation. The main result of the paper is the derivation of coupled evolution equations for the induced electric multipoles  and for the shape functions describing the deformations on the basis of spherical harmonics. Our results, which are consistent with previous models in the appropriate limits, show excellent agreement with fully nonlinear numerical simulations based on an axisymmetric boundary-element formulation and with existing experimental data in the small-deformation regime.
\end{abstract}


\section{Introduction}\label{intro}

Electric fields, when applied to weakly conducting dielectric liquids, can give rise to fluid motions, the study of which forms the field of electrohydrodynamics \citep{melcher1969,saville1997}. In contrast with aqueous electrolytes, ion dissociation in the presence of electric fields is typically negligible in dielectric liquids, so that diffuse Debye layers are absent and fluid motions instead result from the coupling of electric and hydrodynamic stresses acting on interfaces. Electrohydrodynamic phenomena find  widespread industrial applications, such as: inkjet printing \citep{basaran2013,park2007}, electrospraying and atomization of liquids \citep{taylor1964,taylor1969,castellanos2014}, solvent extraction \citep{scott1989}, electrohydrodynamic pumps \citep{laser2004}, and fiber electrospinning  \citep{huang2003}, among others. 

We focus here on the simple problem of electrohydrodynamic deformations of an uncharged leaky dielectric drop suspended in an infinite weakly conducting fluid medium and subject to a steady uniform  electric field. This problem, first studied by \citet{wilson1925}, was originally analyzed under the premise that normal electric stresses acting on an uncharged interface are responsible for deformations \citep{konski1953,harris1957}. Normal stresses, however, can only result in prolate deformations, while experiments have been known to show both prolate and oblate shapes depending on material properties \citep{allan1962}. This paradox was resolved in the pioneering work of \citet{taylor1966}, who recognized that dielectric liquids, while poor conductors, still carry some free charges, which upon application of the field accumulate at the liquid-liquid interface in the form of a surface charge distribution due to the mismatch in electrical properties. Taylor realized that the existence of this surface charge can then give rise to tangential stresses that drive circulatory toroidal currents inside the drop, now known as Taylor vortices. Taylor's theory was able to predict both oblate and prolate shapes and showed good agreement with experiments in  weak  fields. 

Having discovered the importance of surface charge and its contribution to tangential stresses on the interface, \citet{melcher1969} developed a complete framework for studying the electrohydrodynamics of leaky dielectric drops. 
The central result of their model is a surface charge conservation equation that prescribes a balance between transient charge relaxation, the jump in normal Ohmic currents arising from the weak but finite conductivities of the two media, and charge convection on the drop surface by the interfacial fluid velocity. The original model of \cite{taylor1966}, however, neglected transient effects and charge convection and only accounted for first-order deformations in the limit of vanishing electric capillary number $Ca_E$, which compares the magnitude of electric stresses to surface tension. As a result, agreement with experiments was limited to very small deformations, and a number of more detailed theories have been proposed over the years to improve upon this. First, \citet{ajayi1978} extended Taylor's theory by retaining terms to second order in capillary number, but also neglected transients and charge convection. His results, quite surprisingly, showed worse agreement with experiments than the simpler model of Taylor in the case of oblate drops, which is a consequence of the latter approximation. 

Including charge convection, however, is quite challenging as it couples the charge distribution to the resulting fluid flow in a nonlinear fashion. A few computational studies considered its effects \citep{feng1999,supeene2008,lanauze2015} and showed that convection tends to increase deformation in the case of prolate drops but decrease it for oblate drops. On the theoretical side,   \cite{shutov2002} and \cite{shkadov2002} attempted to include it in a small-deformation theory; however, these authors neglected convection at first order and only included it at second order, which as we will show below is incorrect. Very recently, \citet{bandopadhyay2016} studied the dynamics of a drop sedimenting under gravity while subject to an electric field using double asymptotic expansions in electric capillary number $Ca_E$ and electric Reynolds number $Re_E$, which compares electric to viscous stresses. Their theory included linearized charge convection but was limited to small $Re_E$, even though small deviations from drop sphericity only necessitate $Ca_E$ to be small as we show in this work. 

Transient dynamics were also addressed in a few models by including temporal derivatives of shape modes, first by \citet{moriya1986} for perfectly conducting drops, followed by \citet{esmaeeli2011} for weakly conducting drops. The latter theory predicted a monotonic drop deformation leading to the steady drop shape predicted by \cite{taylor1966}. Yet, both experiments \citep{lanauze2015} and numerical simulations \citep{haywood1991,supeene2008} show non-monotonic deformations in cases leading to steady oblate shapes, suggesting an inconsistency in the model. This discrepancy was recently resolved by \citet{lanauze2013}, who showed using a small-deformation theory that either transient charge relaxation or fluid acceleration, combined with transient shape deformations, needs to be included in the model to capture the correct behavior. 

In this work, we present an extension to previous small-deformation theories valid to order $O(Ca_E^2)$ that captures unsteady dynamics. The novelty of our model lies in the theoretical formulation for the complete Melcher-Taylor leaky dielectric model, in which we include transient shape deformation, transient charge relaxation and nonlinear charge convection. As we demonstrate by comparison with boundary element simulations and existing experiments, including both transient phenomena is critical in order to capture the correct shape evolution, and accounting for charge convection leads to improved accuracy in the model predictions as the electric field strength increases. We present the governing equations in \S \ref{probform}. Details of the asymptotic theory are provided in \S \ref{sec:solution} and summarized in \S \ref{sec:summary}, and results of the theory are discussed in \S \ref{sec:results}, where we compare them to experiments as well as boundary element simulations based on an algorithm outlined in appendix~A. We conclude and discuss potential extensions of this work in \S \ref{sec:conclusions}. 

\section{Problem formulation}\label{probform}

\begin{figure}
\centering
\includegraphics{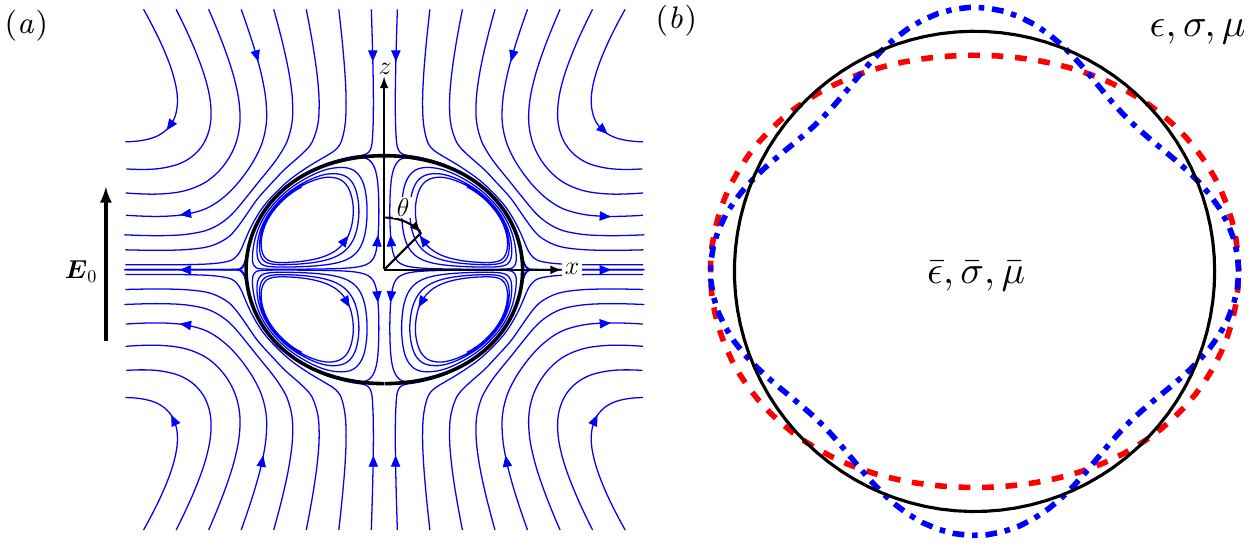}
\caption{(Color online) Problem definition: a liquid drop is placed in a uniform electric field $\boldsymbol{E}_{0}$. ($a$) Spherical coordinates $(r,\theta)$ used in axisymmetric geometry.  Streamlines show the direction of the flow at steady state in the case of an oblately deformed drop. ($b$)  Drop shape expanded on the basis of spherical harmonics. The full line corresponds to the spherical shape, while the dashed line and dash-dotted line correspond to second-order $\mathcal{L}_2$ and fourth-order $\mathcal{L}_{4}$ deformation modes, respectively.}
\label{fig:schematic}
\end{figure}

We analyze the deformation of a neutrally buoyant liquid drop suspended in another liquid and subject to a uniform electric field $\boldsymbol{E}_0=E_0\boldsymbol{\hat{e}}_z$ as shown in figure \ref{fig:schematic}. The drop, with undeformed radius $a$, is assumed to carry no net charge. Both liquids are Newtonian and are treated as leaky dielectrics with constant material properties. The dielectric permittivity, electric conductivity, and dynamic viscosity of the carrying liquid are denoted by ($\epsilon$, $\sigma$, $\mu$), respectively, whereas those of the drop are denoted by  ($\bar\epsilon$, $\bar\sigma$,  $\bar\mu$). The interface $S$ between the two liquids has uniform surface tension $\gamma$ and outward unit normal $\boldsymbol{n}$.

Following the Melcher-Taylor leaky dielectric model \citep{melcher1969}, we assume that any net charge in the system is concentrated at the interface between the two liquids.
Under this condition, the electric potentials $\varphi$ and $\bar\varphi$ outside and inside the drop both satisfy Laplace's equation:
\begin{equation}
\nabla^{2}\varphi=0, \qquad \nabla^2 \bar\varphi=0. \label{eq:laplace}
\end{equation}
The potential is continuous across the interface:
\begin{equation}
\varphi(\boldsymbol{x})=\bar\varphi(\boldsymbol{x}) \qquad \mbox{for} \quad \boldsymbol{x}\in S, \label{eq:continuous}
\end{equation}
and approaches the externally applied potential far away from the drop:
\begin{equation}
\varphi(\boldsymbol{x})\rightarrow \varphi_e(\boldsymbol{x})= -\boldsymbol{E}_0\bcdot\boldsymbol{x} \quad \mbox{as}\quad |\boldsymbol{x}|\rightarrow \infty.  \label{eq:farfield}
\end{equation}
Due to the mismatch in material properties, a surface charge density $q(\boldsymbol{x})$ develops at the interface between the two liquids as the drop polarizes and is given by Gauss's law:
\begin{equation}
q(\boldsymbol{x})=\boldsymbol{n}\bcdot \llbracket \epsilon \boldsymbol{E}(\boldsymbol{x})\rrbracket=\epsilon E^n(\boldsymbol{x})-\bar\epsilon \bar{E}^n(\boldsymbol{x}),   \label{eq:gauss}
\end{equation}
where $\boldsymbol{E}=-\bnabla\varphi$ is the local electric field and $E^n=\boldsymbol{n}\bcdot\boldsymbol{E}$ its normal component. The charge density $q$ evolves due to two distinct mechanisms: Ohmic currents $\boldsymbol{j}=\sigma\boldsymbol{E}$ from the bulk, and surface charge convection by the fluid flow with velocity $\boldsymbol{v}$ along the interface. Accordingly, it satisfies the conservation equation
\begin{equation}
\partial_t q+\boldsymbol{n}\bcdot \llbracket  \boldsymbol{j}\rrbracket +\bnabla_{s}\bcdot (q\boldsymbol{v})=0,
\label{eq:chargeeq}
\end{equation}
where $\bnabla_{s}\equiv ({\mathsfbi{I}}-\boldsymbol{nn})\bcdot\bnabla$ is the surface gradient operator. The flow velocity, which is driven by electric stresses on the interface, satisfies the Stokes equations in both liquids:
\begin{subequations}
\begin{align}
&-\mu \nabla^2\boldsymbol{v}+\bnabla p^H = \boldsymbol{0}, \quad \bnabla\bcdot\boldsymbol{v}=0, \\
&-\bar\mu \nabla^2\bar{\boldsymbol{v}}+\bnabla \bar{p}^H = \boldsymbol{0}, \quad \bnabla\bcdot\bar{\boldsymbol{v}}=0,
\end{align}
\end{subequations}
and is continuous across the interface. Here, $p^H$ denotes the hydrodynamic pressure in the fluid. In the absence of Marangoni effects, the jumps in electric and hydrodynamic tractions balance  interfacial tension forces:
\begin{equation}
\llbracket \boldsymbol{f}^{E}\rrbracket+ \llbracket \boldsymbol{f}^{H}\rrbracket=\gamma (\bnabla_{s}\bcdot\boldsymbol{n})\boldsymbol{n} \quad \mbox{for}\,\,\, \boldsymbol{x}\in S, \label{eq:dynamicBC}
\end{equation}  
where $\bnabla_{s} \bcdot\boldsymbol{n}$ is the total surface curvature.  The jumps in tractions are expressed in terms of the Maxwell stress tensor ${\mathsfbi{T}}^{E}$ and hydrodynamic stress tensor ${\mathsfbi{T}}^{H}$ as
\begin{subequations}
\begin{align}
\llbracket \boldsymbol{f}^{E}\rrbracket&=\boldsymbol{n}\bcdot\llbracket{\mathsfbi{T}}^{E}\rrbracket=\boldsymbol{n}\bcdot\llbracket \epsilon (\boldsymbol{EE}-\tfrac{1}{2}E^{2}{\mathsfbi{I}})\rrbracket, \\
\llbracket \boldsymbol{f}^{H}\rrbracket&=\boldsymbol{n}\bcdot\llbracket{\mathsfbi{T}}^{H}\rrbracket = \boldsymbol{n}\bcdot \llbracket-p^H\,{\mathsfbi{I}}+\mu\left(\bnabla \boldsymbol{v}+\bnabla\boldsymbol{v}^{T}\right)\rrbracket.
\end{align}
\end{subequations}
The jump in electric tractions can be further simplified as
\begin{align}\label{eq:electricf}
\begin{split}
\llbracket\boldsymbol{f}^{E}\rrbracket&=[\epsilon E^{n}-\bar{\epsilon} \bar E^{n}] \boldsymbol{E}^{t}+\tfrac{1}{2}[\epsilon(E^{n2}-E^{t2})-\bar{\epsilon}(\bar E^{n2}-E^{t2})]\boldsymbol{n} \\
&=q\boldsymbol{E}^{t}+\llbracket p^{E}\rrbracket \boldsymbol{n}.
\end{split} 
\end{align}
Here, $\boldsymbol{E}^{t}=(\boldsymbol{\mathsfbi{I}}-\boldsymbol{nn})\bcdot\boldsymbol{E}$ is the tangential electric field, which is continuous across the interface.  The first term on the right hand side captures the tangential electric force arising from the action of the tangential field on the interfacial charge. The second term captures normal electric stresses and can be interpreted as a jump in an electric pressure $p^E$ \citep{lac2007}.

In the remainder of the paper, we scale all lengths by the radius $a$ and times by the Maxwell-Wagner relaxation time $\tau_{MW}$, which is the characteristic time scale for polarization of the drop:
\begin{equation}
\tau_{MW}=\frac{\bar{\epsilon}+ 2\epsilon}{\bar{\sigma}+2\sigma}.
\end{equation}
Electric and hydrodynamic stresses are non-dimensionalized by $\epsilon E_{0}^{2}$ and $\mu/\tau_{MW}$, respectively. Upon scaling of the governing equations, five dimensionless parameters emerge, three of which are ratios of material properties:
\begin{equation}
  Q = \frac{\bar{\epsilon}}{\epsilon}, \quad
  R = \frac{\sigma}{\bar{\sigma}}, \quad
  \lambda = \frac{\bar{\mu}}{\mu}.
  \label{eq1}
\end{equation}
The product $RQ$, which sets the type of deformation and direction of the flow at steady state \citep{lac2007}, can also be interpreted as the ratio of the inner and outer charge relaxation times:
\begin{equation}
RQ=\frac{\bar\tau}{\tau}\qquad \mbox{where}\qquad \tau=\frac{\epsilon}{\sigma}, \quad \bar\tau=\frac{\bar\epsilon}{\bar\sigma}.
\end{equation}

The two remaining dimensionless parameters are chosen as the electric capillary number $Ca_E$ denoting the ratio of electric to capillary forces, and the Mason number $Ma$ denoting the ratio of viscous to electric forces:
\begin{equation}
  Ca_E = \frac{a\epsilon E_0^2}{\gamma}, \qquad
  Ma = \frac{2\mu}{\epsilon \tau_{MW} E_0^2}
 \label{eq:2.11}.
\end{equation}
The Mason number is directly related to the electric Reynolds number $Re_E$ \citep{melcher1969,salipante2010,lanauze2015} as: 
\begin{equation} 
Re_E = \frac{1}{Ma} \frac{2(1+2R)}{R(Q+2)}. \label{eq:2.12} \vspace{-0.1cm}
\end{equation}

\section{Problem solution by domain perturbation\label{sec:solution}}

We solve the governing equations for axisymmetric shapes in the limit of small deformations \citep{taylor1966,ajayi1978,rallison1984}, which occurs when surface tension is strong enough to overcome deformations due to electric stresses. This corresponds to the limit of $Ca_E\rightarrow 0$, and allows us to use an asymptotic approach in which we expand the drop deformation about the spherical shape and all the field variables in a small shape parameter $\delta$ whose relation with $Ca_E$ we explain later.  \vspace{-0.1cm}

\subsection{Shape parametrization and expansion \label{sec:parametrization}}

In axisymmetric geometry, we parametrize the drop shape as a curve $\xi(r,\eta)=0$, where $r=|\boldsymbol{x}|$ is the distance from the drop center and $\eta=\cos\theta$ is the cosine of the polar angle $\theta\in[0,\pi]$ measured from the field direction. For small deviations from sphericity, the drop shape is expanded on the basis of spherical harmonics as 
\begin{equation}
\xi(r,\eta)=r- (1 + \delta f_1 + \delta^2 f_2) + O(\delta^3). \label{eq:3.1}  
\end{equation}
The first- and second-order shape functions are linear combinations of Legendre polymonials $\mathcal{L}_n$ of order $n$: 
\begin{subequations}
\begin{align}
&f_1=f_{12} \mathcal{L}_{2}(\eta), \label{eq:f1} \\
&f_2=f_{20}+f_{22}\mathcal{L}_2(\eta)+f_{24}\mathcal{L}_4(\eta), \label{eq:f2} 
\end{align}
\end{subequations}
where the deformations corresponding to $\mathcal{L}_{2}$ and $\mathcal{L}_4$ are illustrated in figure~\ref{fig:schematic}($b$). We note the orthogonality condition 
\begin{align}
\int^{\uppi}_{0} \mathcal{L}_i(\eta) \mathcal{L}_j (\eta) \sin\theta \,\mathrm{d}\theta = \frac{2}{2i+1} \delta_{ij}, 
\end{align}
which will become useful later. The choice of Legendre functions in equations (\ref{eq:f1})--(\ref{eq:f2}) is a consequence of the quadratic nature of the Maxwell electric stresses acting on the fluid-drop interface, which in a uniform electric field only excite shape modes of order $2^n $ ($n\in \mathbb{Z}^+$). In equations (\ref{eq:f1})--(\ref{eq:f2}) and in the rest of the paper, pairs of indices in coefficients of the form $f_{ij}$ refer to the order $i$ in the small deformation expansion and to the order $j$ of the Legendre polynomial they multiply, respectively. In equation (\ref{eq:f2}), the constant term $f_{20}$ is added to the second-order shape function $f_2$ to negate the perturbation in the drop volume due to the first-order shape function $f_1$: \vspace{-0.1cm}
\begin{equation}
 2\uppi\int^{\uppi}_{0} \int^{r}_{0} \rho^2 \sin\theta \:\mathrm{d}\rho\:\mathrm{d}\theta = \frac{4\uppi}{3} + 4\uppi \delta^2 \left(\frac{f_{12}^2}{5} + f_{20}\right) + O(\delta^3). \vspace{-0.1cm}
\end{equation}
Requiring  terms of order $\delta^2$ to vanish, we get $f_{20} = -f_{12}^2/5$. The outward unit normal, tangent vector and curvature of the interface are also obtained as \citep{ajayi1978}
\begin{subequations} \label{eq:3.2}
\begin{align}
&\boldsymbol{n} = \boldsymbol{\hat{e}}_r - \delta\, \partial_\theta f_1\,\boldsymbol{\hat{e}}_\theta + {O}(\delta^2), \quad \boldsymbol{t} = \boldsymbol{\hat{e}}_\theta + \delta\, \partial_\theta f_1 \,\boldsymbol{\hat{e}}_r + {O}(\delta^2), \label{eq:ntvectors} \\
&\bnabla_{s} \bcdot \boldsymbol{n} = 2 - \delta L[f_1] - \delta^2 \{L[f_2] - 2 f_1(L[f_1] - f_1)\} + {O}(\delta^3), \label{eq:3.2c}
\end{align}
\end{subequations}
where the differential operator $L$ is defined as $L[f] = \partial_{\eta}\{(1-\eta^2) \partial_\eta f\} + 2f$.

Using the above parametrization, the normal and tangential components of any vector $\boldsymbol{v}$ and second-order tensor $\mathsfbi{T}$ on the drop surface are related to their components in spherical coordinates by
\begin{subequations}\label{eq:3.7}
\begin{align}
&v^n=v^r_0+\delta(v_1^r+f_1 \partial_rv_0^r-\partial_\theta f_1v_0^\theta)+O(\delta^2), \label{eq:vnexpansion} \\
&v^t=v^\theta_0+\delta(v_1^\theta+f_1\partial_rv_0^\theta +\partial_\theta f_1v_0^r)+O(\delta^2), \label{eq:vtexpansion} \\
&T^{nn}=T^{rr}_0+\delta(T^{rr}_1+f_1 \partial_rT^{rr}_0-2\partial_\theta f_1T^{r\theta}_0)+O(\delta^2), \label{eq:Tnn} \\
&T^{nt}=T^{r\theta}_0+\delta[T^{r\theta}_1+f_1 \partial_rT^{r\theta}_0+ \partial_\theta f_1(T^{rr}_0-T^{\theta\theta}_0)]+O(\delta^2), \label{eq:Ttt}
\end{align}
\end{subequations}
where the terms on the right-hand side are to be evaluated at $r=1$. These expressions will be useful below in determining the electric field, fluid velocity and stress distributions on the drop surface.

\subsection{Electric problem}

\subsubsection{Spherical harmonic expansion}

We first present the solution to the electric problem, which consists in solving equations (\ref{eq:laplace})--(\ref{eq:farfield}) asymptotically. The electric potential outside and inside the drop can be expanded in powers of $\delta$ as
\begin{subequations} \label{eq:3.4}
\begin{align}
\varphi &= \varphi_e(r,\theta) + \varphi_0 (r,\theta) + \delta \varphi_1 (r,\theta) + {O}(\delta^2), \label{eq:3.4a} \\
\bar{\varphi} &= \varphi_e(r,\theta) + \bar{\varphi}_0 (r,\theta) + \delta \bar{\varphi}_1 (r,\theta) +  {O}(\delta^2), \label{eq:3.4b}
\end{align}
\end{subequations}
which automatically satisfies the far-field boundary condition (\ref{eq:farfield}).  We have yet to enforce continuity of the potential across the interface. To this end, we employ a domain perturbation approach in which all the boundary conditions are enforced approximately on the undeformed spherical surface $r=1$. The potential on the interface is first expanded in the neighborhood of $r=1$ using Taylor series:
\begin{subequations}\label{eq:3.6}
\begin{align}
\varphi &= \varphi_e + \varphi_0 + \delta \left[ \varphi_1 +  f_1\partial_r(\varphi_e + \varphi_0) \right] +  {O}(\delta^2), \label{eq:3.6a} \\
\bar{\varphi} &= \bar{\varphi}_e + \bar{\varphi}_0 + \delta \left[ \bar{\varphi}_1 +  f_1 \partial_r(\varphi_e + \bar{\varphi}_0) \right] +  {O}(\delta^2). \label{eq:3.6b}
\end{align}
\end{subequations}
Applying continuity (\ref{eq:continuous}) and matching terms of zeroth- and first-order in $\delta$ provides two boundary conditions at $r=1$:
\begin{subequations}
\begin{align}
\varphi_0 &= \bar{\varphi}_0, \label{eq:0continuity}\\
\varphi_1 +  f_1 \partial_r(\varphi_e + \varphi_0) &= \bar{\varphi}_1 +  f_1 \partial_r(\varphi_e +  \bar{\varphi}_0). \label{eq:1continuity}
\end{align}
\end{subequations}
The zeroth-order problem, which is identical to the case of a sphere, is easily solved using decaying and growing spherical harmonics in terms of electric dipoles $P_{01}$, $\bar{P}_{01}$:
\begin{subequations} \label{eq:3.9}
\begin{align}
\varphi_0 &= {P_{01}}r^{-2} \mathcal{L}_1 (\eta), \label{eq:3.9a}\\
\bar{\varphi}_0 &= {\bar{P}_{01} r} \mathcal{L}_1(\eta), \label{eq:3.9b}
\end{align}
\end{subequations}
and we require that $\bar{P}_{01}=P_{01}$ to satisfy (\ref{eq:0continuity}); solving for $P_{01}$ will require application of the charge conservation equation (\ref{eq:chargeeq}) as detailed below. After substitution into equation (\ref{eq:1continuity}), we obtain a new first-order boundary condition:
\begin{equation} 
\varphi_1 - \bar{\varphi}_1 = 3 f_1 P_{01} \mathcal{L}_1(\eta)=3f_{12} P_{01} \mathcal{L}_1(\eta)\mathcal{L}_2(\eta)=\tfrac{3}{5}f_{12} P_{01}[2\mathcal{L}_1(\eta)+3\mathcal{L}_3(\eta)]. \label{eq:11continuity}
\end{equation}
The order of the polynomials appearing on the right-hand side suggests representing the first-order potentials in terms of both dipoles $P_{11}$, $\bar P_{11}$ and octupoles $P_{13}$, $\bar P_{13}$:
\begin{subequations}\label{eq:3.13}
\begin{align}
& \varphi_1 = P_{11} r^{-2}\mathcal{L}_1(\eta)+P_{13} r^{-4}\mathcal{L}_3(\eta), \label{eq:3.13a} \\
& \bar{\varphi}_1 = \bar{P}_{11} r \mathcal{L}_1(\eta)+\bar{P}_{13} r^3 \mathcal{L}_3(\eta), \label{eq:3.13b}
\end{align}
\end{subequations}
and application of the boundary condition (\ref{eq:11continuity}) yields the relations
\begin{equation}
\bar{P}_{11}=P_{11} -\tfrac{6}{5} f_{12}P_{01}, \qquad \bar{P}_{13}=P_{13}-\tfrac{9}{5}f_{12}P_{01}.
\end{equation}

Having determined the electric potential, we can also obtain asymptotic expressions for the normal and tangential electric fields $E^n=-\boldsymbol{n}\bcdot\bnabla\varphi$ and $E^t=-\boldsymbol{t}\bcdot\bnabla\varphi$ on the drop surface. Applying equation (\ref{eq:vnexpansion}), we find
\begin{equation}
E^n=E^n_0+\delta E^n_1+O(\delta^2)=E^n_{01}\mathcal{L}_{1}(\eta)+\delta [E^n_{11}\mathcal{L}_{1}(\eta)+E^n_{13}\mathcal{L}_3(\eta)]+O(\delta^2), \vspace{-0.1cm}
\end{equation}
with a similar expansion for $\bar{E}^n$, with coefficients\vspace{-0.1cm}
\begin{subequations}\label{eq:3.17}
\begin{align}
&E_{01}^n = 1 + 2P_{01}, && \bar{E}_{01}^n = 1 - P_{01}, \label{eq:3.17a} \\
&E_{11}^n = 2P_{11} - \tfrac{6}{5}f_{12}(1+P_{01}), && \bar{E}_{11}^n = -P_{11} - \tfrac{6}{5}f_{12}(1 -2P_{01}), \label{eq:3.17b} \\
&E_{13}^n = 4P_{13} + \tfrac{6}{5}f_{12} (1-4P_{01}), && \bar{E}_{13}^n = -3P_{13} + \tfrac{6}{5} f_{12} \left(1 + \tfrac{7}{2}P_{01} \right). \label{eq:3.17c}\vspace{-0.1cm}
\end{align}
\end{subequations}
Finally, the expansion for the tangential electric field, which is continuous across the interface, is obtained using equation (\ref{eq:vtexpansion}) and is written
\begin{equation}
E^t=E^t_0+\delta E^t_1+O(\delta^2)=E^t_{00}\sin \theta + \delta [E^t_{10}+E^t_{12}\mathcal{L}_{2}(\eta)]\sin\theta+O(\delta^2),\vspace{-0.1cm}
\end{equation}
where\vspace{-0.1cm}
\begin{subequations}\label{eq:3.21}
\begin{align}
E_{00}^t &= -(1-P_{01}), \label{eq:3.21a}\\
E_{10}^t &= P_{13} + P_{11} -f_{12}(1+2P_{01}), \label{eq:3.21b}\\
E_{12}^t &= 5P_{13} -f_{12}(2+7P_{01}). \label{eq:3.21c}\vspace{-0.1cm}
\end{align}
\end{subequations}

\subsubsection{Charge conservation and moment equations}

To complete the solution of the electric problem, equations must be derived for the moments $P_{01}$, $P_{11}$ and $P_{13}$, which are time-dependent. These can be obtained as ordinary differential equations by application of the charge conservation equation (\ref{eq:chargeeq}). First, we expand the charge density in powers of $\delta$ as
\begin{equation}
q=q_0+\delta q_1 + O(\delta^2)=q_{01}\mathcal{L}_1(\eta) + \delta [q_{11}\mathcal{L}_1(\eta)+q_{13}\mathcal{L}_3(\eta)]+O(\delta^2), \label{eq:chargeexp}
\end{equation}
where the coefficients are obtained using Gauss's law as
\begin{subequations}\label{eq:3.22}
\begin{align}
q_{01} & = E^n_{01}-Q \bar{E}^n_{01}= 1+2P_{01} - Q(1-P_{01}), \label{eq:3.22a}\\
q_{11} & = E^n_{11}-Q \bar{E}^n_{11}=2P_{11} - \tfrac{6}{5}f_{12}(1+P_{01})-Q[-P_{11} - \tfrac{6}{5}f_{12}(1 -2P_{01})], \label{eq:3.22b}\\
q_{13} & = E^n_{13}-Q \bar{E}^n_{13}= 4P_{13} + \tfrac{6}{5}f_{12} (1-4P_{01})-Q[-3P_{13} + \tfrac{6}{5} f_{12} \left(1 + \tfrac{7}{2}P_{01} \right)]. \label{eq:3.22c}
\end{align}
\end{subequations}
Similarly, we expand the jump in Ohmic currents $\boldsymbol{n}\bcdot\llbracket\boldsymbol{j}\rrbracket=\llbracket j \rrbracket^n$, scaled here by $\bar \sigma E_0$, as
\begin{equation}
\llbracket {j}\rrbracket^n = \llbracket {j}\rrbracket^n_{0}+\delta \llbracket {j}\rrbracket^n_{1}+O(\delta^2)= \llbracket {j}\rrbracket^n_{01}\mathcal{L}_1(\eta) + \delta \{  \llbracket {j}\rrbracket^n_{11}\mathcal{L}_1(\eta)+\llbracket {j}\rrbracket^n_{13}\mathcal{L}_3(\eta) \}+O(\delta^2), \label{eq:jexp}
\end{equation}
where Ohm's law provides
\begin{subequations}\label{eq:3.23}
\begin{align}
\llbracket j \rrbracket^n_{01}& = R E^{n}_{01}-\bar{E}^n_{01}=R(1+2P_{01}) - 1+P_{01}, \label{eq:3.23a}\\
\llbracket j \rrbracket^n_{11}& = RE^{n}_{11}-\bar{E}^n_{11}=R[2P_{11} - \tfrac{6}{5}f_{12}(1+P_{01})]+P_{11} + \tfrac{6}{5}f_{12}(1 -2P_{01}), \label{eq:3.23b}\\
\llbracket j \rrbracket^n_{13}& = RE^{n}_{13}-\bar{E}^n_{13}= R[4P_{13} + \tfrac{6}{5}f_{12} (1-4P_{01})]+3P_{13} - \tfrac{6}{5} f_{12} \left(1 + \tfrac{7}{2}P_{01} \right) . \label{eq:3.23c}
\end{align}
\end{subequations}
Finally, we formally expand the charge convection term in equation (\ref{eq:chargeeq})  as
\begin{align}
\begin{split}
\bnabla_s&\bcdot(q\boldsymbol{v})=[\bnabla_s\bcdot(q\boldsymbol{v})]_0+\delta [\bnabla_s\bcdot(q\boldsymbol{v})]_1+O(\delta^2),  \\
&=[\bnabla_s\bcdot(q\boldsymbol{v})]_{01}\mathcal{L}_1(\eta)+\delta \{[\bnabla_s\bcdot(q\boldsymbol{v})]_{11}\mathcal{L}_1(\eta)+[\bnabla_s\bcdot(q\boldsymbol{v})]_{13}\mathcal{L}_3(\eta) \}+O(\delta^2)
\end{split} \label{eq:convexp}
\end{align}
where we have introduced the Legendre coefficients
\begin{align}
[\bnabla_s\bcdot(q\boldsymbol{v})]_{ij} =\frac{2j+1}{2}\int_{0}^{\uppi} [\bnabla_s\bcdot(q\boldsymbol{v})]_{i} \mathcal{L}_{j}(\eta) \sin\theta \,\mathrm{d}\theta. \label{eq:Legendrecoefs}
\end{align}
Detailed expressions for these coefficients require knowledge of the interfacial velocity $\boldsymbol{v}$, whose calculation is presented in \S \ref{sec:flow}. 

Substituting the expansions (\ref{eq:chargeexp}), (\ref{eq:jexp}) and (\ref{eq:convexp}) into the charge conservation equation (\ref{eq:chargeeq}), matching powers of $\delta$, and applying orthogonality of Legendre polynomials leads to a set of relaxation equations for the charge coefficients. In dimensionless form, these read
\begin{equation}
\dot q_{ij}+\frac{Q+2}{1+2R}\,\llbracket j \rrbracket ^n_{ij}+[\bnabla_s\bcdot(q\boldsymbol{v})]_{ij}=0,
\end{equation}
where the dot in the first term denotes differentiation with respect to time.
If we further express   $q_{ij}$ and $\llbracket j \rrbracket^n_{ij}$ in terms of $P_{01}$, $P_{11}$ and $P_{13}$ using (\ref{eq:3.22}) and (\ref{eq:3.23}), we arrive at a set of hierarchical differential equations for the dipole and octupole moments:
\begin{align} 
&\dot P_{01}+P_{01}=\frac{1-R}{1+2R}-\frac{1}{Q+2}[\bnabla_s\bcdot(q\boldsymbol{v})]_{01}, \label{eq:momenteqs01}  \\
\begin{split}&\dot P_{11}+P_{11}=\frac{\mathrm{d}}{\mathrm{d}t}\left[\frac{6}{5}f_{12}\left(P_{01} \frac{1+2Q}{2+Q} + \frac{1-Q}{2+Q} \right)\right]+\frac{6}{5}f_{12}\left(P_{01}\frac{R+2}{2R+1} - \frac{1-R}{2R+1} \right)\\
&\qquad\qquad\,\,\,\,\,\,-\frac{1}{Q+2}[\bnabla_s\bcdot(q\boldsymbol{v})]_{11},  \label{eq:momenteqs11}
\end{split}  \\
\begin{split}&\dot P_{13}+\frac{Q+2}{3Q+4}\frac{4R+3}{2R+1}P_{13}=\frac{\mathrm{d}}{\mathrm{d}t} \left[\frac{6}{5}f_{12}\left(P_{01} \frac{8+7Q}{8+6Q} - \frac{1-Q}{4+3Q}\right)\right] \\
&\qquad\qquad\,\,\,\,\,\, +\frac{6}{5}f_{12}\frac{Q+2}{3Q+4}\left(P_{01} \frac{8R+7}{4R+2} + \frac{1-R}{2R+1} \right)-\frac{1}{3Q+4}[\bnabla_s\bcdot(q\boldsymbol{v})]_{13}.\end{split}  \label{eq:momenteqs13}
\end{align}
These coupled ordinary differential equations constitute the main result of this section. The external forcing in these equations is encapsulated in the first term on the right-hand side of (\ref{eq:momenteqs01}), which describes the effect of the applied electric field on the leading-order dipole moment. Solving (\ref{eq:momenteqs01})--(\ref{eq:momenteqs13}) requires the Legendre coefficients of the charge convection term  as well as the first-order shape coefficient $f_{12}$. These unknowns will be determined below after we solve for the fluid flow, which affects both interfacial charge convection and droplet deformation. 

\subsection{Flow problem: streamfunction formulation\label{sec:flow}}

We now turn to the calculation of the fluid flow outside and inside the drop. Upon application of the field, electric stresses develop at the interface leading to deformations and flow. Since the flow is axisymmetric, we use a Stokes streamfunction $\Psi(r,\theta)$ to determine the fluid velocity, which has components
\begin{equation}
v^r=\frac{1}{r^2\sin\theta}\partial_{\theta}\Psi, \qquad v^\theta=-\frac{1}{r\sin\theta}\partial_{r}\Psi, \label{eq:velcomponents}
\end{equation} 
in spherical coordinates. 
The streamfunction satisfies the biharmonic equation $\nabla^4 \Psi=0$, the general solutions to which outside and inside the drop are \citep{kim2013}:
\begin{align}
\Psi &= \sum_{n=2}^\infty (A_nr^{-n+1} + B_n r^{-n+3}) \mathcal{G}_n(\eta), \qquad \bar{\Psi} = \sum_{n=2}^\infty (\bar{A}_n r^n + \bar{B}_n r^{n+2} ) \mathcal{G}_n(\eta),
\end{align}
where $\mathcal{G}_n(\eta)$ are Gegenbauer functions of degree $-1/2$ of the first kind \citep{abramowitz72}. They are related to Legendre polynomials and are regular everywhere in $-1 \leq \eta \leq 1$:
\begin{align}
\mathcal{G}_n(\eta) = \frac{\mathcal{L}_{n-2}(\eta)-\mathcal{L}_n(\eta)}{2n-1}, ~~~ n \geq 2.
\end{align}
The first two functions are defined as $\mathcal{G}_0(\eta)=1$ and $\mathcal{G}_1(\eta)=-\eta$, and we also note the property: $\mathcal{G}_{n}'(\eta)=-\mathcal{L}_{n-1}(\eta)$.

Following the same methodology as for the electric problem, we seek solutions as expansions in powers of $\delta$.
As will become evident in \S \ref{sec:stressbalance} when performing the stress balance on the interface, the zeroth- and first-order electric stresses acting on the interface at most induce fluid motions of the form
\begin{subequations}
\begin{align}
\Psi &= \Psi_{03}\mathcal{G}_{3}(\eta) + \delta [\Psi_{13}\mathcal{G}_{3}(\eta)+\Psi_{15}\mathcal{G}_{5}(\eta)] + {O}(\delta^2), \\
\bar\Psi &= \bar\Psi_{03}\mathcal{G}_{3}(\eta) + \delta [\bar\Psi_{13}\mathcal{G}_{3}(\eta)+\bar\Psi_{15}\mathcal{G}_{5}(\eta)] + {O}(\delta^2),
\end{align}
\end{subequations}
where 
\begin{subequations}
\begin{align}
& \Psi_{03}= A_{03}r^{-2} + B_{03}, && \bar{\Psi}_{03} = \bar{A}_{03}r^3 + \bar{B}_{03}r^5, \\
& \Psi_{13}=A_{13}r^{-2} + B_{13}, && \bar{\Psi}_{13} = \bar{A}_{13}r^3 + \bar{B}_{13}r^5, \\
& \Psi_{15} = A_{15} r^{-4} + B_{15} r^{-2}, && \bar{\Psi}_{15} = \bar{A}_{15} r^5 + \bar{B}_{15} r^{7}.
\end{align}
\end{subequations}
In particular, the flow is entirely determined by twelve coefficients that are functions of time and that we proceed to solve for by application of the boundary conditions. 

\subsection{Kinematic boundary condition}

The kinematic boundary condition relates the shape deformation to the fluid velocity so as to satisfy the no-slip and no-penetration boundary conditions at the interface. The streamfunction $\Psi$ can be used to determine the normal and tangential components of the fluid velocity on the drop surface, which are obtained by combining equations (\ref{eq:vnexpansion})--(\ref{eq:vtexpansion}) and (\ref{eq:velcomponents}) as 
\begin{subequations}\label{eq:3.55}
\begin{align}
v^n &= v^n_{02} \mathcal{L}_2(\eta) + \delta [v^n_{10} + v^n_{12} \mathcal{L}_2(\eta) + v^n_{14} \mathcal{L}_4(\eta)] + {O}(\delta^2), \\
\bar{v}^n &= \bar{v}^n_{02} \mathcal{L}_2(\eta) + \delta [\bar{v}^n_{10} + \bar{v}^n_{12} \mathcal{L}_2(\eta) + \bar{v}^n_{14} \mathcal{L}_4(\eta)] + {O}(\delta^2), \\
v^t &= v^t_{01} \mathcal{L}_1(\eta)\sin\theta + \delta [v^t_{11} \mathcal{L}_1(\eta) + v^t_{13} \mathcal{L}_3(\eta)]\sin\theta + {O}(\delta^2), \\
\bar{v}^t &= \bar{v}^t_{01} \mathcal{L}_1(\eta)\sin\theta + \delta [\bar{v}^t_{11} \mathcal{L}_1(\eta) + \bar{v}^t_{13} \mathcal{L}_3(\eta)]\sin\theta + {O}(\delta^2).
\end{align}
\end{subequations}
The zeroth-order coefficients are found to be 
\begin{subequations}
\begin{align}
&v^n_{02} = A_{03} + B_{03} , && \bar{v}^n_{02} = \bar{A}_{03} + \bar{B}_{03}, \\
&v^t_{01} = A_{03}, && \bar{v}^t_{01} = -\tfrac{3}{2}\bar{A}_{03} - \tfrac{5}{2}\bar{B}_{03}.
\end{align}
\end{subequations}
At first order, they read
\begin{subequations}\label{eq:3.58}
\begin{align}
&v^n_{10} = -\tfrac{2}{5}f_{12}(A_{03}+B_{03} ), \\
&v^n_{12} = A_{13} + B_{13}  -\tfrac{2}{7}f_{12}(3A_{03}+2B_{03}),\\
&v^n_{14} = A_{15} + B_{15} -\tfrac{12}{35}f_{12}(8A_{03}+3B_{03}),   \\
&\bar{v}^n_{10} = -\tfrac{2}{5}f_{12}(\bar{A}_{03} + \bar{B}_{03}), \\
&\bar{v}^n_{12} = \bar{A}_{13}+\bar{B}_{13} - \tfrac{1}{7}f_{12}(\bar{A}_{03} - \bar{B}_{03}), \\
&\bar{v}^n_{14} = \bar{A}_{15}+\bar{B}_{15} + \tfrac{6}{35}f_{12}(9\bar{A}_{03} + 19\bar{B}_{03}),
\end{align}
\end{subequations}
whereas those of the tangential velocity are given by
\begin{subequations}\label{eq:3.59}
\begin{align}
&v^t_{11} = A_{13} + \tfrac{3}{5}A_{15}+\tfrac{3}{10}B_{15}-\tfrac{2}{5}f_{12} (7A_{03}+3B_{03}),\\
&v^t_{13} = \tfrac{7}{5}A_{15}+\tfrac{7}{10}B_{15}-\tfrac{3}{5}f_{12} (7A_{03}+3B_{03}), \\
&\bar{v}^t_{11} = -\tfrac{3}{2}\bar{A}_{13}  -\tfrac{5}{2}\bar{B}_{13}-\tfrac{3}{4}\bar{A}_{15} - \tfrac{21}{20}\bar{B}_{15}-\tfrac{3}{5}f_{12}(3\bar{A}_{03} + 7\bar{B}_{03} ),\\
&\bar{v}^t_{13}=-\tfrac{7}{4}\bar{A}_{15} - \tfrac{49}{20}\bar{B}_{15}-\tfrac{9}{10}f_{12}(3\bar{A}_{03} + 7\bar{B}_{03} ).
\end{align}
\end{subequations}
The no-penetration boundary condition is expressed as $v^n=\bar{v}^n=\dot{\xi}$, which provides the four relations
\refstepcounter{equation}
$$
v^n_{02}=\bar{v}^n_{02} = \delta \dot{f}_{12}, \qquad \qquad v^n_{10}=\bar{v}^n_{10} = \delta \dot{f}_{20}, \eqno{(\theequation{\mathit{a},\mathit{b}})} \label{eq:nopen1}
$$
$$
v^n_{12}=\bar{v}^n_{12} = \delta \dot{f}_{22},  \qquad \qquad v^n_{14}=\bar{v}^n_{14} =\delta \dot{f}_{24}. \eqno{(\theequation{\mathit{c},\mathit{d}})}
$$
Similarly, the no-slip boundary condition $v^t=\bar{v}^t$ dictates that
\refstepcounter{equation}
$$
v_{01}^t=\bar{v}_{01}^t, \qquad  v_{11}^t=\bar{v}_{11}^t, \qquad   v_{13}^t=\bar{v}_{13}^t \eqno{(\theequation{\mathit{a}, \mathit{b},\mathit{c}})}. \label{eq:noslip}
$$
The matching of orders in equation~(\ref{eq:nopen1}) might seem surprising at first due to the presence of terms involving $\delta$ on the right-hand side. However, it is the only possible solution as the leading-order term in $\dot{\xi}$ involves $\delta$. This implies that temporal derivatives of the shape functions in fact scale as $\delta^{-1}$, suggesting that the characteristic time scale for the shape transient is not the Maxwell-Wagner relaxation used here for non-dimensionalization. This point will be made clearer in \S \ref{sec:firstorder}.

The zeroth-order boundary conditions (\ref{eq:nopen1}$a$) and (\ref{eq:noslip}$a$) provide us with the relations
\begin{align}
A_{03}=-B_{03}+\delta\dot{f}_{12},\quad
\bar{A}_{03}=-B_{03}+\tfrac{7}{2}\delta\dot{f}_{12},\quad
\bar{B}_{03}=B_{03}-\tfrac{5}{2}\delta\dot{f}_{12}. \label{eq:zeroBC}
\end{align}
Using these relations together with the condition that $f_{20}=-f_{12}^2/5$ obtained in \S \ref{sec:parametrization} from volume conservation, it is easy to show that (\ref{eq:nopen1}$b$) is trivially satisfied. The remaining first-order boundary conditions then yield six additional equations that can be combined to show that
\begin{subequations}\label{eq:oneBC}
\begin{align}
&A_{13}=-B_{13}-\tfrac{2}{7}f_{12}B_{03}+\delta\dot{f}_{22}+\tfrac{6}{7}\delta f_{12}\dot{f}_{12},\\
&\bar{A}_{13}=-B_{13}+\tfrac{3}{7}f_{12}B_{03}+\tfrac{7}{2}\delta\dot{f}_{22}+\tfrac{1}{2}\delta f_{12}\dot{f}_{12},\\
&\bar{B}_{13}=B_{13}-\tfrac{5}{7}f_{12}B_{03}-\tfrac{5}{2}\delta\dot{f}_{22}+\tfrac{5}{14}\delta f_{12}\dot{f}_{12},\\
&A_{15}=-B_{15}-\tfrac{12}{7}f_{12}B_{03}+\delta\dot{f}_{24}+\tfrac{96}{35}\delta f_{12}\dot{f}_{12},\\
&\bar{A}_{15}=-B_{15}-\tfrac{6}{7}f_{12}B_{03}+\tfrac{11}{2}\delta\dot{f}_{24}+\tfrac{3}{35}\delta f_{12}\dot{f}_{12},\\
&\bar{B}_{15}=B_{15}-\tfrac{6}{7}f_{12}B_{03}-\tfrac{9}{2}\delta\dot{f}_{24}+\tfrac{93}{35}\delta f_{12}\dot{f}_{12}. 
\end{align}
\end{subequations}
Equations (\ref{eq:zeroBC})--(\ref{eq:oneBC}) therefore allow us to reduce the number of flow unknowns to three, namely $B_{03}$, $B_{13}$ and $B_{15}$.

\subsection{Dynamic boundary condition\label{sec:stressbalance}}

We now proceed to enforce the dynamic boundary condition of equation (\ref{eq:dynamicBC}), which in dimensionless form reads
\begin{align}
 \boldsymbol{n} \bcdot\llbracket \boldsymbol{T}^E  \rrbracket + \frac{Ma}{2} \boldsymbol{n} \bcdot \llbracket \boldsymbol{T}^H \rrbracket = \frac{1}{Ca_E} (\bnabla_s \bcdot \boldsymbol{n}) \boldsymbol{n}, \label{eq:stressbalance}
\end{align}
and requires us to evaluate electric and hydrodynamic stresses on the interface. 

\subsubsection{Electric stress}

As previously shown in equation (\ref{eq:electricf}), the jump in electric tractions can be decomposed into tangential and normal components, both of which involve quadratic products of expansions derived above. The tangential component $q\boldsymbol{E}^t=qE^t\boldsymbol{t}$ is continuous and is expanded as
\begin{equation}
qE^t=[qE^t]_{01}\mathcal{L}_{1}(\eta)\sin\theta+\delta\{ [qE^t]_{11}\mathcal{L}_{1}(\eta)+[qE^t]_{13}\mathcal{L}_{3}(\eta) \}\sin\theta+O(\delta^2),
\end{equation}
with coefficients
\begin{subequations}
\begin{align}
&[qE^t]_{01}=q_{01}E_{01}^t, \\
&[qE^t]_{11}=q_{01}E_{11}^t + \tfrac{2}{5}q_{01}E_{13}^t + q_{11}E_{01}^t, \\
&[qE^t]_{13}=\tfrac{3}{5} q_{01}E_{13}^t + q_{13}E_{01}^t,
\end{align}
\end{subequations}
where the various products on the right-hand side are  easily evaluated using equations (\ref{eq:3.21}) and (\ref{eq:3.22}). Similarly, the expansion for the jump in electric pressure in equation (\ref{eq:electricf}) is found to be
\begin{equation}
\llbracket p^E\rrbracket = \llbracket p^E\rrbracket _{00}+\llbracket p^E\rrbracket _{02}\mathcal{L}_{2}(\eta)+ \delta \{ \llbracket p^E\rrbracket _{10}+ \llbracket p^E\rrbracket _{12}\mathcal{L}_{2}(\eta)+\llbracket p^E\rrbracket _{14}\mathcal{L}_{4}(\eta) \} + O(\delta^2),
\end{equation}
where the coefficients are obtained as 
\begin{subequations}
\begin{align}
&\llbracket p^E\rrbracket _{00}=\tfrac{1}{6}(E_{01}^{n2}-Q\bar{E}_{01}^{n2})+\tfrac{1}{3}(Q-1)E_{00}^{t2},\\
&\llbracket p^E\rrbracket_{02}=\tfrac{1}{3}(E_{01}^{n2}-Q\bar{E}_{01}^{n2})-\tfrac{1}{3}(Q-1)E_{00}^{t2},\\
&\llbracket p^E\rrbracket_{10}=\tfrac{1}{3}(E_{01}^nE_{11}^n-Q\bar{E}_{01}^n\bar{E}_{11}^n)+\tfrac{2}{3}E^t_{00}(E_{10}^t-\tfrac{1}{5}E_{12}^t),\\
\begin{split} &\llbracket p^E\rrbracket_{12}=\tfrac{2}{3}(E_{01}^nE_{11}^n-Q\bar{E}_{01}^n\bar{E}_{11}^n)+\tfrac{3}{7}(E_{01}^nE_{13}^n-Q\bar{E}_{01}^n\bar{E}_{13}^n) \\
&\qquad\qquad+\tfrac{2}{3}E_{00}^t(\tfrac{5}{7}E_{12}^t-E_{10}^t),\end{split} \\
&\llbracket p^E\rrbracket_{14}=\tfrac{4}{7}(E_{01}^nE_{13}^n-Q\bar{E}_{01}^n\bar{E}_{13}^n)-\tfrac{12}{35}E_{00}^tE_{12}^t,
\end{align}
\end{subequations}
and can be calculated using equations (\ref{eq:3.17}) and (\ref{eq:3.21}).

\subsubsection{Hydrodynamic stress}

The jump in hydrodynamic tractions is evaluated using equations (\ref{eq:Tnn})--(\ref{eq:Ttt}), in which the requisite components of the stress tensor in spherical coordinates are obtained from the velocity components as
\begin{subequations}
\begin{align}
&T^{H,rr} = -p^H + 2 \partial_r v^r,&& \bar{T}^{H,rr} = -\bar{p}^H + 2\lambda  \partial_r \bar{v}^r,\\
&T^{H,r\theta} = r^{-1} \partial_\theta v^r + r \partial_r(v^\theta r^{-1}), && \bar{T}^{H,r\theta} = \lambda [r^{-1} \partial_\theta \bar{v}^r + r \partial_r(\bar{v}^\theta r^{-1})], \\
& T^{H,\theta\theta}=-p^H+2r^{-1}(\partial_{\theta} v_{\theta}+v_r), && \bar{T}^{H,\theta\theta}=-\bar{p}^H+2r^{-1}\lambda(\partial_{\theta} \bar{v}_{\theta}+\bar{v}_r).
\end{align}
\end{subequations}
The diagonal stress components $T^{H,rr}$ and $T^{H,\theta\theta}$ involve the fluid pressure $p^H$, which can be obtained from the velocity by integration of the momentum equation. After some algebra, the jumps in hydrodynamic stresses induced by the zeroth- and first-order streamfunctions $\Psi_0$, $\Psi_1$, scaled with $\mu/\tau_{MW}$, are found as
\begin{subequations}
\begin{align}
\begin{split}
\llbracket T^H\rrbracket ^{nn}&=\llbracket T^H\rrbracket ^{nn}_0+\delta \llbracket T^H\rrbracket ^{nn}_1+{O}(\delta^2)\\
&=\llbracket T^H\rrbracket ^{nn}_{00}+\llbracket T^H\rrbracket ^{nn}_{02}\mathcal{L}_2(\eta) \\
&\,\,\,\,\,\,+\delta \{ \llbracket T^H\rrbracket ^{nn}_{10}+\llbracket T^H\rrbracket ^{nn}_{12}\mathcal{L}_2(\eta)+ \llbracket T^H\rrbracket ^{nn}_{14}\mathcal{L}_4(\eta) \} +{O}(\delta^2), \end{split} \\
\begin{split}\llbracket T^H\rrbracket^{nt}&=\llbracket T^H\rrbracket^{nt}_0+\delta \llbracket T^H\rrbracket^{nt}_1+{O}(\delta^2)\\
&=\llbracket T^H\rrbracket^{nt}_{01}\mathcal{L}_1(\eta)\sin\theta+\delta\{\llbracket T^H\rrbracket^{nt}_{11} \mathcal{L}_1(\eta)+\llbracket T^H\rrbracket^{nt}_{13}\mathcal{L}_3(\eta)\} \sin\theta+{O}(\delta^2),\end{split}
\end{align}
\end{subequations}
where the various coefficients can all be expressed in terms of $B_{03}$, $B_{13}$, $B_{15}$ after making use of equations (\ref{eq:zeroBC})--(\ref{eq:oneBC}). At zeroth order, we find:
\begin{subequations} \label{eq:Thzero}
\begin{align}
\llbracket T^H\rrbracket_{00}^{nn} &= \llbracket p^H \rrbracket_{00},\\
\llbracket T^H\rrbracket_{02}^{nn} &= (2+3\lambda)B_{03}-\tfrac{1}{2}(16+19\lambda)\delta\dot{f}_{12},\\
\llbracket T^H\rrbracket_{01}^{nt} &= 5(1+\lambda)B_{03}-\tfrac{1}{2}(16+19\lambda)\delta\dot{f}_{12}.
\end{align}
\end{subequations}
Similarly, at first order,
\begin{subequations} \label{eq:Thone}
\begin{align}
\llbracket T^H\rrbracket_{10}^{nn}&=\llbracket p^H \rrbracket_{10}+ \tfrac{2}{5}(-1+11\lambda)B_{03}f_{12}+\tfrac{1}{5}(8-43\lambda)\delta\dot{f}_{12}f_{12},\\
\begin{split} \llbracket T^H\rrbracket_{12}^{nn}&=(2+3\lambda)B_{13}+\tfrac{1}{7}(-8+13\lambda)B_{03}f_{12}-\tfrac{1}{2}(16+19\lambda)\delta\dot{f}_{22} \\
&\,\,\,\,\,\,-\tfrac{105}{14}\lambda\,\delta f_{12}\dot{f}_{12},\end{split} \\
\begin{split} \llbracket T^H\rrbracket_{14}^{nn}&=\tfrac{3}{10}(4+5\lambda)B_{15}+\tfrac{3}{35}(28+37\lambda)B_{03}f_{12}-\tfrac{3}{4}(16+17\lambda)\delta\dot{f}_{24}\\
&\,\,\,\,\,\,-\tfrac{3}{70}(32+47\lambda)\delta f_{12}\dot{f}_{12}, \end{split} \\
\begin{split}\llbracket T^H\rrbracket_{11}^{nt}&=5(1+\lambda)B_{13}+\tfrac{27}{10}(1+\lambda)B_{15}-\tfrac{4}{35}(33+18\lambda)B_{03}f_{12}\\
&\,\,\,\,\,\, -\tfrac{1}{2}(16+19\lambda)\delta\dot{f}_{22}-\tfrac{9}{20}(16+17\lambda)\delta\dot{f}_{24}+\tfrac{2}{175}(227-466\lambda)\delta f_{12}\dot{f}_{12}, \end{split} \\
\begin{split} \llbracket T^H\rrbracket_{13}^{nt}&=\tfrac{63}{10}(1+\lambda)B_{15}-\tfrac{9}{5}(1+\lambda)B_{03}f_{12}-\tfrac{21}{20}(16+17\lambda)\delta\dot{f}_{24} \\
&\,\,\,\,\,\,+\tfrac{9}{50}(4-7\lambda)\delta f_{12}\dot{f}_{12}. \end{split}
\end{align}
\end{subequations}
In equations (\ref{eq:Thzero}) and (\ref{eq:Thone}),  $\llbracket p^H \rrbracket_{00}$ and $\llbracket p^H \rrbracket_{10}$ denote uniform hydrostatic pressure jumps that do no affect drop shape. 

\subsubsection{Stress balance}

The electric and hydrodynamic traction jumps can now be substituted into the stress balance (\ref{eq:stressbalance}) to satisfy the dynamic boundary condition. In the normal direction, the stress balance requires:\vspace{-0.1cm}
\begin{subequations}\label{eq:normalstressbalance}
\begin{align} 
& \llbracket p^E\rrbracket_{00}+\frac{Ma}{2} \llbracket T^H\rrbracket_{00}^{nn} =\frac{2}{Ca_E}, \\
& \llbracket p^E\rrbracket_{02}+\frac{Ma}{2} \llbracket T^H\rrbracket_{02}^{nn} =\frac{4}{Ca_E}\delta f_{12}, \\
& \llbracket p^E\rrbracket_{10}+\frac{Ma}{2} \llbracket T^H\rrbracket_{10}^{nn}= -\frac{2}{Ca_E}\delta f_{12}^2, \\
& \llbracket p^E\rrbracket_{12}+\frac{Ma}{2} \llbracket T^H\rrbracket_{12}^{nn} = \frac{4}{Ca_E} \delta (f_{22} - \tfrac{5}{7}f_{12}^2 ), \\
& \llbracket p^E\rrbracket_{14}+\frac{Ma}{2} \llbracket T^H\rrbracket_{14}^{nn} = \frac{18}{Ca_E} \delta(f_{24} - \tfrac{2}{7} f_{12}^2).
\end{align} \vspace{-0.1cm}
\end{subequations}
In the tangential direction, it yields\vspace{-0.1cm}
\begin{subequations}\label{eq:tanstressbalance}
\begin{align}
& [qE^t]_{01} + \frac{Ma}{2} \llbracket T^H\rrbracket_{01}^{nt} = 0, \\
& [qE^t]_{11} + \frac{Ma}{2} \llbracket T^H\rrbracket_{11}^{nt} = 0, \\
& [qE^t]_{13} + \frac{Ma}{2}\llbracket T^H\rrbracket_{13}^{nt} = 0. 
\end{align}
\end{subequations}
The above balances now allow us to define more explicitly the value of the small deformation parameter $\delta$. The driving force for the flow is the tangential electric stress $q\boldsymbol{E}^t$, which according to equations (\ref{eq:tanstressbalance}) induces hydrodynamic tractions scaling with $O(Ma^{-1})$. The magnitude of the resulting flow therefore is such that all normal tractions, both electric and hydrodynamic, in equation (\ref{eq:normalstressbalance}) are of order $O(1)$. Balancing these tractions with surface tension forces thus requires us to choose $\delta\propto Ca_E$. For consistency with previous small deformation theories, we define $\delta$ explicitly as
\begin{align}
\delta=\frac{3\,Ca_E}{4(1+2R)^2}. \label{eq:delta}
\end{align}
In particular, we find no restriction on the magnitude of the Mason or electric Reynolds numbers, which remain arbitrary in our model.

\subsection{Nonlinear charge convection}

As a final calculation, we determine the Legendre coefficients of the nonlinear convection term in the charge convection equation (\ref{eq:chargeeq}). The convection term is straightforward to calculate after applying the identity 
\begin{equation}
\bnabla_s \bcdot (q\boldsymbol{v}) = q v^n (\bnabla_s \bcdot \boldsymbol{n}) + \bnabla_s \bcdot (q\boldsymbol{v}^t),
\end{equation}
in which the expansions for $q$, $v^n$, $\boldsymbol{v}^t=v^t\boldsymbol{t}$, and $\bnabla_s \bcdot \boldsymbol{n}$ can be substituted together with 
\begin{equation}
\bnabla_s= [\mathsfbi{I} - \boldsymbol{\hat{e}}_r \boldsymbol{\hat{e}}_r + \delta (\boldsymbol{\hat{e}}_r \boldsymbol{\hat{e}}_\theta +\boldsymbol{\hat{e}}_\theta \boldsymbol{\hat{e}}_r )]\bcdot\bnabla +O(\delta^2).
\end{equation}
All calculations done, the relevant Legendre coefficients appearing in equations (\ref{eq:momenteqs01})--(\ref{eq:momenteqs13}) for the dipole and octupole moments are found to be
\begin{subequations}
\begin{align}
&[\bnabla_s \bcdot (q \boldsymbol{v})]_{01}  = - \tfrac{2}{5}q_{01}B_{03}+ \tfrac{6}{5} q_{01} \delta\dot{f}_{12} , \\
\begin{split}& [\bnabla_s \bcdot (q \boldsymbol{v})]_{11} =  \tfrac{2}{5}q_{01} A_{13} + \tfrac{2}{5}q_{11} A_{03} - \tfrac{6}{35} q_{13}A_{03}-\tfrac{54}{35}q_{01}A_{03}f_{12} \\
&\qquad\qquad\qquad\,\,\, +\tfrac{4}{5}q_{01}\delta \dot{f}_{22}+\tfrac{4}{5}q_{11}\delta \dot{f}_{12}+\tfrac{18}{35}q_{13}\delta \dot{f}_{12}+\tfrac{38}{35}q_{01}\delta f_{12}\dot{f}_{12}, \end{split} \\
\begin{split} & [\bnabla_s \bcdot (q \boldsymbol{v})]_{13} =  \tfrac{8}{5}q_{01}A_{13} +  \tfrac{4}{3} q_{01}A_{15} +\tfrac{2}{3} q_{01}B_{15}+ \tfrac{8}{5}q_{11} A_{03} \\
&\qquad\qquad\qquad\,\,\, + \tfrac{4}{15}q_{13}A_{03}  -\tfrac{104}{15}q_{01}A_{03}f_{12}+\tfrac{6}{5}q_{01}\delta \dot{f}_{22}+\tfrac{8}{9}q_{01}\delta \dot{f}_{24} \\
&\qquad\qquad\qquad\,\,\, +\tfrac{6}{5}q_{11}\delta\dot{f}_{12}+\tfrac{8}{15} q_{13}\delta\dot{f}_{12}-\tfrac{4}{5}q_{01}\delta f_{12} \dot{f}_{12}. \end{split}
\end{align}
\end{subequations}

\section{Summary of the small-deformation theory\label{sec:summary}}

The set of asymptotic expansions obtained in \S \ref{sec:solution} provides a closed system of equations for all unknown coefficients. We summarize here the results of the theory and outline the solution procedure at first and second order. We also compare and contrast our predictions with the existing theories of \cite{taylor1966}, \cite{ajayi1978}, \cite{esmaeeli2011} and \cite{lanauze2013}. \vspace{-0.05cm}

\subsection{Taylor deformation parameter}

For easy comparison with previous theories and experiments, we introduce Taylor's deformation parameter $\mathcal{D}$, defined as \vspace{-0.15cm}
\begin{equation}
\mathcal{D}=\frac{r^+-r^-}{r^++r^-}, \vspace{-0.05cm}
\end{equation}
where $r^+$ and $r^-$ denote the longest and shortest distances of any point on the interface from the drop center, respectively. The sign of $\mathcal{D}$ distinguishes between  oblate ($\mathcal{D}<0$) and prolate ($\mathcal{D}>0$)  shapes.   For an axisymmetric drop, $r^+$ and $r^-$ are reached at $\theta=0$ and $\uppi/2$, respectively:  \vspace{-0.05cm}
\begin{subequations} 
\begin{align}
&r^+=r(0) = 1 + \delta f_{12} + \delta^2\left(f_{20} + f_{22} + f_{24}\right)+O(\delta^3), \\
&r^- =r(\uppi/2)= 1 - \tfrac{1}{2}\delta f_{12} + \delta^2\left(f_{20} - \tfrac{1}{2}f_{22} + \tfrac{3}{8} f_{24}\right)+O(\delta^3), \vspace{-0.1cm}
\end{align}
\end{subequations} 
from which we find \vspace{-0.1cm}
\begin{align}
\mathcal{D}= \tfrac{3}{4} \left[\delta f_{12} + \delta^2 \left(f_{22} + \tfrac{5}{12} f_{24} - \tfrac{1}{4}f_{12}^2\right) \right] + {O}(\delta^3). \label{eq:Taylorparameter}
\end{align} \vspace{-0.3cm}

\subsection{First-order theory\label{sec:firstorder}}

We first summarize the first-order theory, which allows us to compare our results with those of \cite{taylor1966}, \cite{esmaeeli2011} and \cite{lanauze2013}. The zeroth-order stress balance equations (\ref{eq:normalstressbalance}$b$) and (\ref{eq:tanstressbalance}$a$), together with the dipole relaxation equation (\ref{eq:momenteqs01}), provide three coupled equations for the three unknowns $B_{03}$, ${f}_{12}$ and ${P}_{01}$. We first eliminate $B_{03}$ by combining (\ref{eq:normalstressbalance}$b$) and (\ref{eq:tanstressbalance}$a$), and after manipulations we arrive at a coupled system of first-order ordinary differential equations of the form \vspace{-0.05cm}
\begin{equation}
\frac{\mathrm{d}}{\mathrm{d}t}\left[
\begin{tabular}{c}
$P_{01}$ \\
$f_{12}$
\end{tabular}\right]=\mathcal{F}_1(P_{01},f_{12};Ca_E,Ma,R,Q,\lambda), \label{eq:firstordertheory} \vspace{-0.05cm}
\end{equation}
where $\mathcal{F}_1$ is a nonlinear function whose explicit form is cumbersome and is omitted here for brevity. 
These equations can be integrated numerically in time subject to initial conditions. In all of the results shown below, we assume that the drop surface is initially spherical and does not carry any charge at $t=0$, which provides the initial conditions: \vspace{-0.1cm}
\begin{equation}
 P_{01}(0)=\frac{Q-1}{Q+2}, \qquad f_{12}(0)=0. \vspace{-0.05cm} \label{eq:initial}
\end{equation}

Equations (\ref{eq:firstordertheory}) can easily be compared to previous first-order theories. First, neglecting charge convection decouples the dipole evolution equation  from the fluid problem, yielding the simple relaxation equation  \vspace{-0.05cm}
\begin{align}
\dot{P}_{01} + P_{01} = \frac{1-R}{1+2R} , \vspace{-0.1cm}
\end{align}
the solution to which is:\vspace{-0.1cm} 
\begin{align}
P_{01} =  \frac{1-R}{1+2R} + \left(\frac{Q-1}{Q+2} - \frac{1-R}{1+2R} \right)\mathrm{e}^{-t}. \label{eq:P01sol} \vspace{-0.1cm}
\end{align}
Substituting (\ref{eq:P01sol}) into equation (\ref{eq:firstordertheory}) then yields a simplified model which is similar to that of \cite{lanauze2013} when the effect of fluid inertia is negligible. 
If we further neglect charge relaxation, we can easily solve for the transient deformation parameter as
\begin{equation}
\mathcal{D}(t)=\mathcal{D}_{T}(1-\mathrm{e}^{-t/\tau_d}) \qquad \mbox{where}\qquad  \tau_d = \frac{a\mu}{\gamma}\frac{(19\lambda+16)(2\lambda+3)}{40(\lambda+1)}, \label{eq:Dt1storder}
\end{equation}
which matches the result of \citet{esmaeeli2011}. In particular, the viscous-capillary time scale $\tau_d$ emerges as the characteristic time scale for shape deformations, which also rationalizes the seeming contradiction in the matching of terms in the kinematic boundary of equation~(\ref{eq:nopen1}). Here, $\mathcal{D}_{T}$ is the steady first-order deformation parameter first obtained by \cite{taylor1966} as
\begin{equation}
\mathcal{D}_T  = \frac{9}{16} \frac{\Phi_T}{(1+2R)^2} Ca_E 
\end{equation}
in terms of Taylor's discriminating function $\Phi_T$: 
\begin{equation}
\Phi_T = (1-R)^2 + R(1-RQ) \left[2 + \frac{3}{5}\frac{2+3\lambda}{1+\lambda} \right]. \label{eq:2.13} 
\end{equation}
Note that equation (\ref{eq:Dt1storder}) predicts an exponential relaxation towards the steady drop shape and therefore fails to capture the non-monotonic transient deformation observed in experiments and simulations \citep{lanauze2015} and also predicted by the full solution of equations  (\ref{eq:firstordertheory}) as we discuss in \S \ref{sec:results}.

\subsection{Second-order theory}

The first-order theory can then be improved by solution of the second-order equations, which involve the additional unknowns $B_{13}$, $B_{15}$, $f_{22}$, $f_{24}$, $P_{11}$, and $P_{13}$. These are provided by the first-order normal and tangential stress balances of equations (\ref{eq:normalstressbalance}$c$), (\ref{eq:normalstressbalance}$e$), (\ref{eq:tanstressbalance}$b$) and (\ref{eq:tanstressbalance}$c$), together with the moment evolution equations (\ref{eq:momenteqs11})--(\ref{eq:momenteqs13}). The flow unknowns $B_{13}$ and $B_{15}$ can be eliminated by manipulating equations (\ref{eq:normalstressbalance}$c$) and (\ref{eq:tanstressbalance}$b$) for $B_{13}$, and equations (\ref{eq:normalstressbalance}$e$) and (\ref{eq:tanstressbalance}$c$) for $B_{15}$. When combined with the moment evolution equations, this yields a system a coupled differential equations of the form
\begin{equation}
\frac{\mathrm{d}}{\mathrm{d}t}\left[
\begin{tabular}{c}
$P_{11}$ \\
$P_{13}$ \\
$f_{22}$ \\
$f_{24}$
\end{tabular}\right]=\mathcal{F}_2(P_{11},P_{13},f_{22},f_{24};P_{01},f_{12};Ca_E,Ma,R,Q,\lambda), \label{eq:2ndordertheory}
\end{equation}
where $\mathcal{F}_{2}$ is another nonlinear function. Once again, these equations can be integrated in time numerically to obtain the multipole moments as well as shape functions entering Taylor's deformation parameter of equation (\ref{eq:Taylorparameter}). The initial conditions for these variables in the case of an initially spherical and uncharged drop are
\begin{equation}
P_{11}(0)=P_{13}(0)=f_{22}(0)=f_{24}(0)=0.
\end{equation}
If charge convection is neglected, equations (\ref{eq:momenteqs11})--(\ref{eq:momenteqs13}) for the moments become uncoupled from the flow problem and only involve electric parameters. At steady state, the first-order multipole moments are then obtained as
\begin{align}
P_{11} = \frac{6}{5}f_{12} \left(\frac{1-R}{1+2R}\right)^2,\qquad
P_{13} = \frac{9}{5}f_{12} \frac{1-R}{1+2R},
\end{align}
which matches equations (25) and (26) in the work of \citet{ajayi1978}. The numerical codes solving systems (\ref{eq:firstordertheory}) and (\ref{eq:2ndordertheory}) are available upon request.

\section{Results and discussion \label{sec:results}}

\begin{table}
  \begin{center}
\def~{\hphantom{0}}
  \begin{tabular}{cccccccccc}
     System & $\epsilon/\epsilon_0$ & $\bar{\epsilon}/\epsilon_0$ & $\sigma$ & $\bar{\sigma}$ & $\mu$ & $\bar{\mu}$ & $\gamma$  & $a$ &$E_0$ \\[1pt]
      & & & (S.$\text{m}^{-1}$) & (S.$\text{m}^{-1}$)& (Pa.s)& (Pa.s) & (mN.$\text{m}^{-1}$) & (mm) & (kV.$\text{cm}^{-1}$) \\  \hline
     1a & 4.9 & 2.8 & $5.8 \times 10^{-11}$  & $0.2 \times 10^{-11}$ & 0.68 & 0.05 & 4.5 & 2.0 & 1.6 \\[1pt]
     1b & 4.9 & 2.8 & $5.8 \times 10^{-11}$  & $0.2 \times 10^{-11}$ & 0.68 & 0.05 & 4.5 & 2.0 & 2.1 \\[1pt]
     1c & 4.9 & 2.8 & $5.8 \times 10^{-11}$  & $0.2 \times 10^{-11}$ & 0.68 & 0.05 & 4.5 & 2.0 & 6.1 \\[1pt]
     2a & 5.3 & 3.0 & $4.5 \times 10^{-11}$  & $0.12 \times 10^{-11}$ & 0.69 & 0.97 & 4.5 & 1.4 & 0.45--2.0 \\[1pt]
     2b & 5.3 & 3.0 & $4.5 \times 10^{-11}$  & $0.12 \times 10^{-11}$ & 0.69 & 0.97 & 4.5 & 4.2 & 0.26--1.2 \\ \hline 
  \end{tabular}
  \caption{Material properties: systems 1 and  2 correspond to the experiments of \citet{lanauze2015} and \citet{salipante2010}, respectively. $\epsilon_0=8.8542 \times 10^{-12}\,\text{F.m}^{-1}$ denotes the permittivity of vacuum.} \label{table:dimensional}
  \end{center}
\end{table}

We now compare our theoretical results with existing experimental data, previous small-deformation theories, as well as full nonlinear numerical simulations using an axisymmetric boundary element method described in appendix A. The material properties, drop sizes and electric field strengths are chosen as in table \ref{table:dimensional} to match the experimental values of \cite{lanauze2015} (system 1), who measured transient drop dynamics, and of \cite{salipante2010} (system 2) for steady deformations, and corresponding dimensionless parameter values are provided in table \ref{table:dimensionless}. Both of these studies considered oblate drops. We also present a few results on prolate drops, for which we use the experimental values of \cite{ha2000a} (system 3). Their study, however, did not report all the material properties required to construct all five dimensionless parameters in our model; we choose to set the values of the electric capillary number and Mason number to $Ca_E=0.3$ and $Ma=1$ in this case.

\begin{table}
  \begin{center}
\def~{\hphantom{0}}
  \begin{tabular}{ccccccc}
     System & $R$ & $Q$ & $\lambda$ & $Ca_E$ & $Ma$ \\ \hline
     1a & 29.0 & 0.57 & 0.074 & 0.49 & 1.30 \\[1pt]
     1b & 29.0 & 0.57 & 0.074 & 0.85 & 0.75 \\[1pt]
     1c & 29.0 & 0.57 & 0.074 & 7.18 & 0.09 \\[1pt]
     2a & 37.5 & 0.57 & 1.41 & 0.03--0.6 & 0.54--10.8 \\[1pt]
     2b & 37.5 & 0.57 & 1.41 & 0.03--0.6 & 1.6--32\\[1pt]
     3 & 0.1 & 1.37 & 1 & 0.3 & 1\\ \hline
  \end{tabular}
  \caption{Dimensionless parameters corresponding to the material properties of table 1:  systems 1, 2 and 3 correspond to the experiments of \citet{lanauze2015}, \citet{salipante2010} and \citet{ha2000a}, respectively.}
  \label{table:dimensionless}
  \end{center}
\end{table}

\subsection{Effect of transient charge relaxation and shape deformation}

\begin{figure}
\centering
\includegraphics[width=13.1cm]{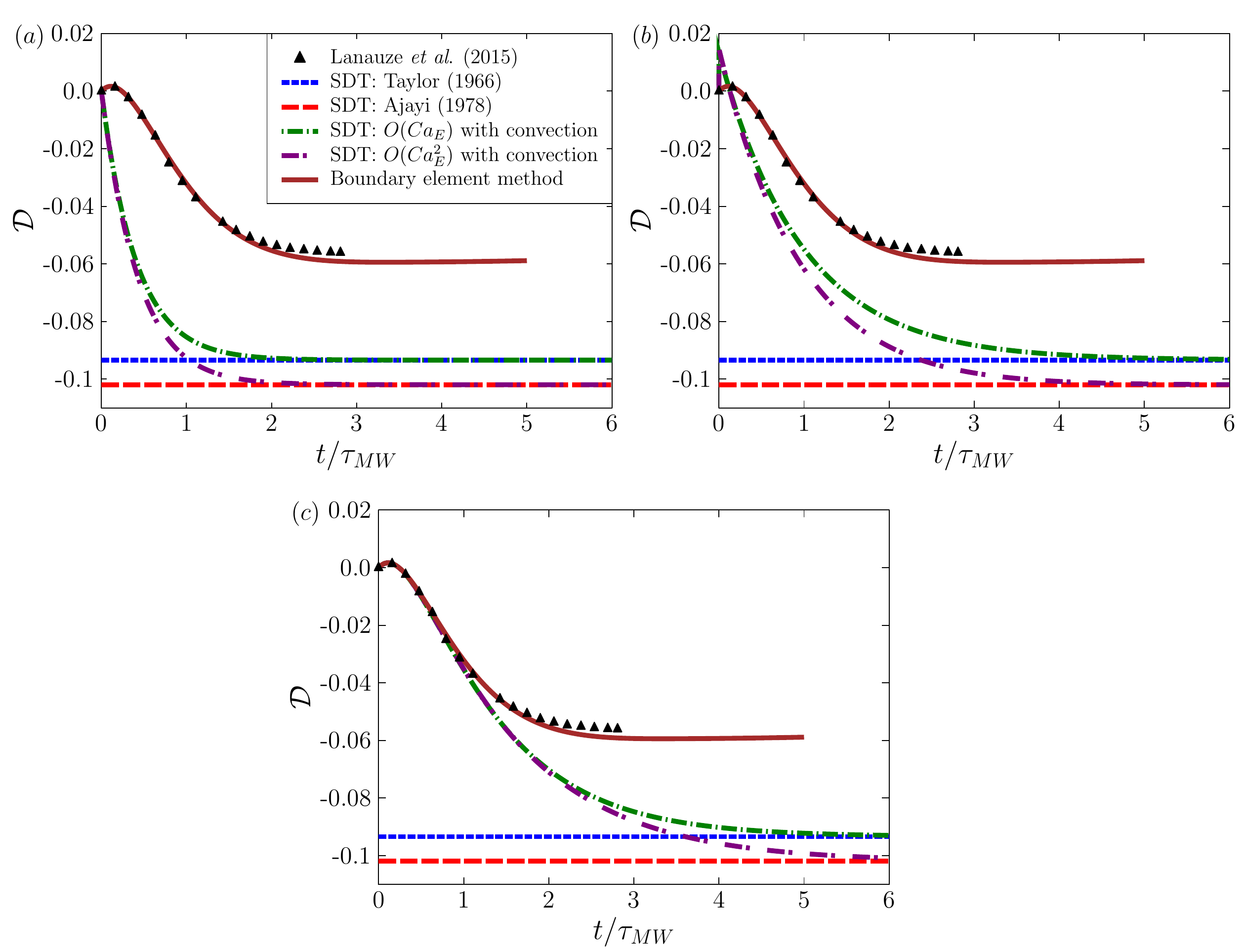}
\caption{(color online) Deformation parameter $\mathcal{D}$ as a function of time for the parameters of system 1b in the absence of charge convection: ($a$) effect of transient charge relaxation only (no transient shape relaxation), ($b$) effect of transient shape relaxation (no transient charge relaxation), and ($c$) effect of both transient shape and charge relaxation. Symbols show experimental data of \citet{lanauze2015}. Boundary element simulation results using the full nonlinear model and the algorithm of appendix~A are also shown.}
\label{fig:transientchargeshape}
\end{figure}

In this section, we first neglect nonlinear charge convection and focus on the effects of transient charge relaxation and transient shape deformation alone. Here we adopt the experimental values of system 1b. The drop deformation is plotted as a function of time in figure~\ref{fig:transientchargeshape} for three distinct cases. In figure~\ref{fig:transientchargeshape}($a$), both nonlinear charge convection and transient charge relaxation are neglected. In this case, the only time-dependence enters through the temporal derivatives of the shape functions. We find that the drop shape become oblate ($\mathcal{D}<0$), and our theoretical results asymptote at long times towards the steady-state predictions of \citet{taylor1966} and \citet{ajayi1978} at first- and second-order, respectively. Both steady states, however, overpredict the drop deformation, and it is found, rather curiously, that the theory performs more poorly at second order than at first order; this was already noted by \citet{ajayi1978} and is a consequence of neglecting charge convection as further discussed below. The transient is also poorly captured: the model predicts a monotonic increase of the drop deformation towards the oblate steady state and fails to capture the initial dynamics seen in experiments, where the drop first adopts a prolate shape before becoming oblate. Figure~\ref{fig:transientchargeshape}($b$) shows the opposite situation in which transient shape deformation is neglected but transient charge relaxation is included. In this case, the shape instantaneously adjusts to the charge distribution, which explains the immediate deformation to a prolate shape at $t=0$ as a result of the instantaneous polarization of the drop according to equation~(\ref{eq:initial}). The deformation subsequently relaxes monotonically towards its steady oblate value. However, accounting for both transient phenomena in figure~\ref{fig:transientchargeshape}($c$) captures the transient dynamics correctly while still evolving towards the steady deformation values of \citet{taylor1966} and \citet{ajayi1978} in the absence of charge convection. These results underscore the importance of including all transient effects in the model if one wants to capture the correct shape dynamics. 

\subsection{Effect of nonlinear charge convection}

\begin{figure}
\centering
\includegraphics[width=13.1cm]{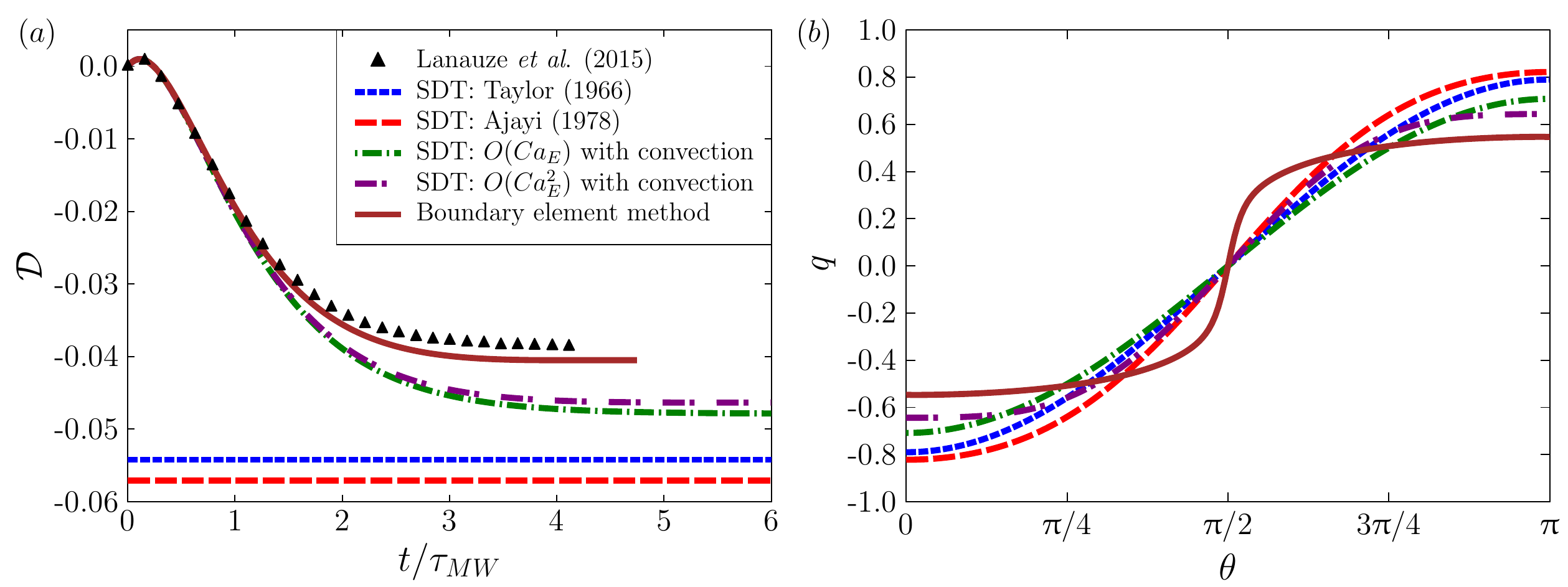}
\caption{(color online) ($a$) Deformation parameter $\mathcal{D}$ as a function of time for the parameters of system 1a. ($b$) Steady interfacial charge profile. The plots show experimental results of \citet{lanauze2015}, fully nonlinear boundary element simulations, first- and second-order small-deformation theory (with nonlinear charge convection), and the steady results of  \citet{taylor1966} and \citet{ajayi1978} that neglected charge convection.}
\label{fig:1a}
\end{figure}

We now turn to the full theoretical model, which includes transient charge and shape relaxation as well as nonlinear charge convection. As we show here, the main effect of charge convection is to reduce the strength of the interfacial velocity, thereby causing oblate drops to deform less but prolate drops to deform more in agreement with computational studies \citep{feng1999,lanauze2015}. We first consider the dynamics in a relatively weak electric field using the parameters of system 1a in figure~\ref{fig:1a}. First, we note in figure~\ref{fig:1a}($a$) that the boundary element simulations perform best and capture both the transient and the steady state with very good accuracy. Our small deformation theory with charge convection also captures the transient very well but still slightly overpredicts the steady deformation parameter, albeit not as much as the models of \cite{taylor1966} and \citet{ajayi1978}. Interestingly, we find that while the second-order theory of Ajayi is worse than the first-order theory of Taylor in the absence of charge convection, such is not the case in our model where including second-order terms is seen to improve the solution. The poor performance of Ajayi's model is a direct consequence of the neglect of charge convection, which results in a stronger dipole moment and in turn leads to larger deformations. Charge convection by the flow, however, causes the transport of positive and negative charges from the poles towards the equator, thus effectively reducing the induced dipole. This point is evident in figure~\ref{fig:1a}($b$) showing the steady charge distribution on the drop surface, where we see that the second-order theory with charge convection best approximates the charge profile from boundary element simulations. This numerical charge profile, however, exhibits a sharper transition from negative to positive values at the equator.

\begin{figure}
\centering
\includegraphics[width=13.1cm]{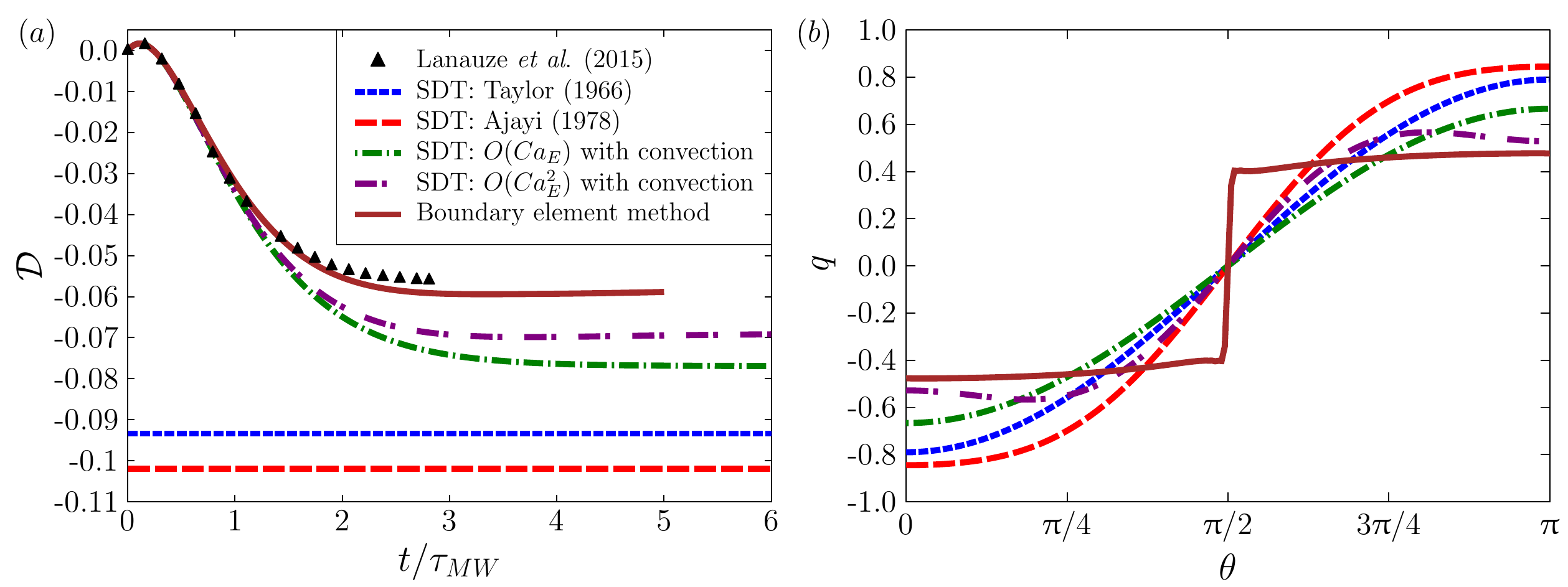}
\caption{(color online) ($a$) Deformation parameter $\mathcal{D}$ as a function of time for the parameters of system 1b. ($b$) Steady interfacial charge profile. For these parameter values, the charge distribution predicted by the boundary element simulation develops a discontinuity at the equator. See supplementary online materials for a movie showing the dynamics and flow field in this case. }
\label{fig:1b}
\end{figure}

\begin{figure}
\centering
\includegraphics[width=13.1cm]{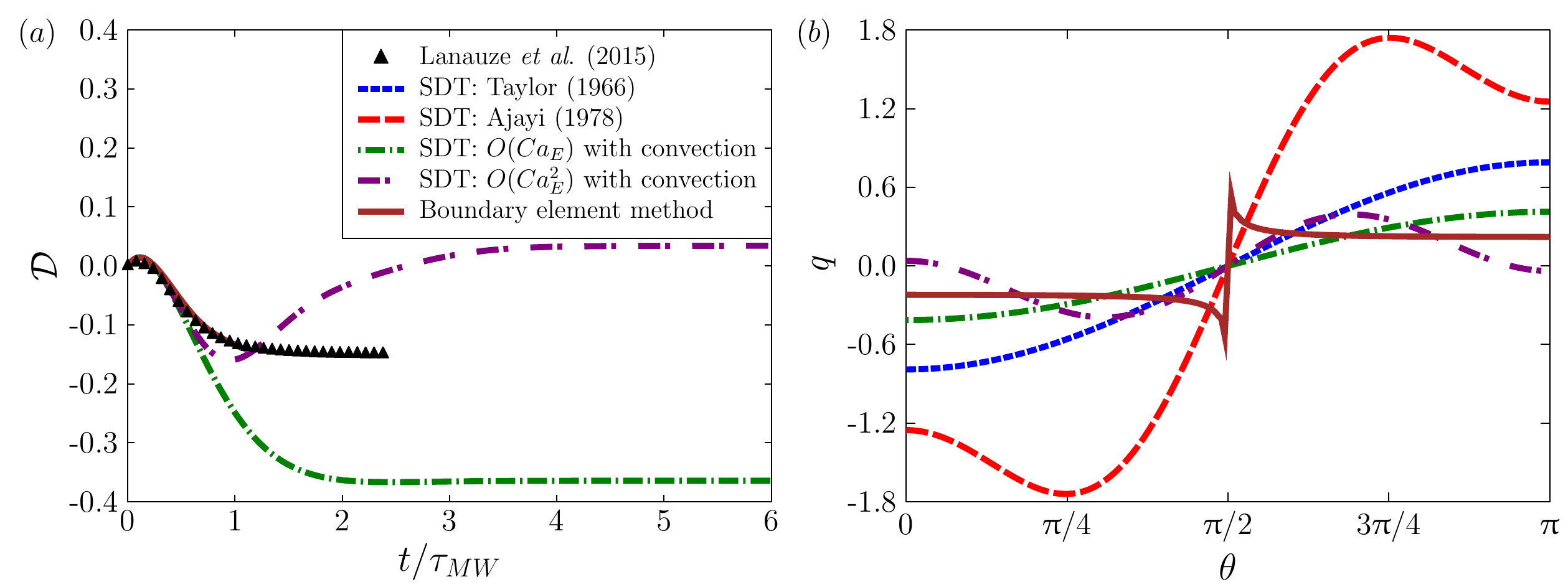}
\caption{(color online) ($a$) Deformation parameter $\mathcal{D}$ as a function of time for the parameters of system 1c. ($b$) Steady interfacial charge profile. The steady deformation values predicted by the models of \cite{taylor1966} and \citet{ajayi1978} in this case are $-0.75$ and $-1.40$, respectively, and out of the frame of figure ($a$). For these parameter values, the charge discontinuity at the equator is so severe that the boundary element simulations blow up before reaching steady state; in this case, the charge profile shown in ($b$) corresponds to a time before the instability develops.  }
\label{fig:1c}
\end{figure}

The effect of increasing field strength is shown in figure~\ref{fig:1b} corresponding to system 1b. Unsurprisingly, stronger fields cause larger drop deformations, which are not as easily captured by the theory. While the boundary element simulation matches the experimental data quite well, our nonlinear small-deformation theory captures the transient well but shows a significant departure at steady state. Nevertheless, the second-order theory still outperforms all previous theoretical models. The difficulty in capturing the steady state accurately can be understood by considering the charge profile in figure~\ref{fig:1b}($b$), where a sharp gradient is observed across the equator in the numerical data from boundary element simulations. This sharp gradient cannot be captured using only two Legendre functions as in the expansion of equation (\ref{eq:chargeexp}), which explains the discrepancy. The problem becomes yet more severe in stronger fields, as shown in figure~\ref{fig:1c} in the case of system 1c. There, an actual discontinuity appears in the charge profile, leading to the very poor performance of small-deformation theories and to numerical instabilities in the boundary element simulation, which blows up before reaching steady state. The formation of a charge shock in strong fields was first observed in the simulations of \cite{lanauze2015}, who also were not able to resolve it numerically using their boundary element algorithm based on spline interpolation. The boundary element method used here and described in appendix A solves the charge conservation equation using finite volumes and yet is still unable to capture the discontinuity, suggesting that higher-order non-oscillating numerical schemes should be employed towards this purpose \citep{leveque2002}.

\begin{figure}
\centering
\includegraphics[width=13.1cm]{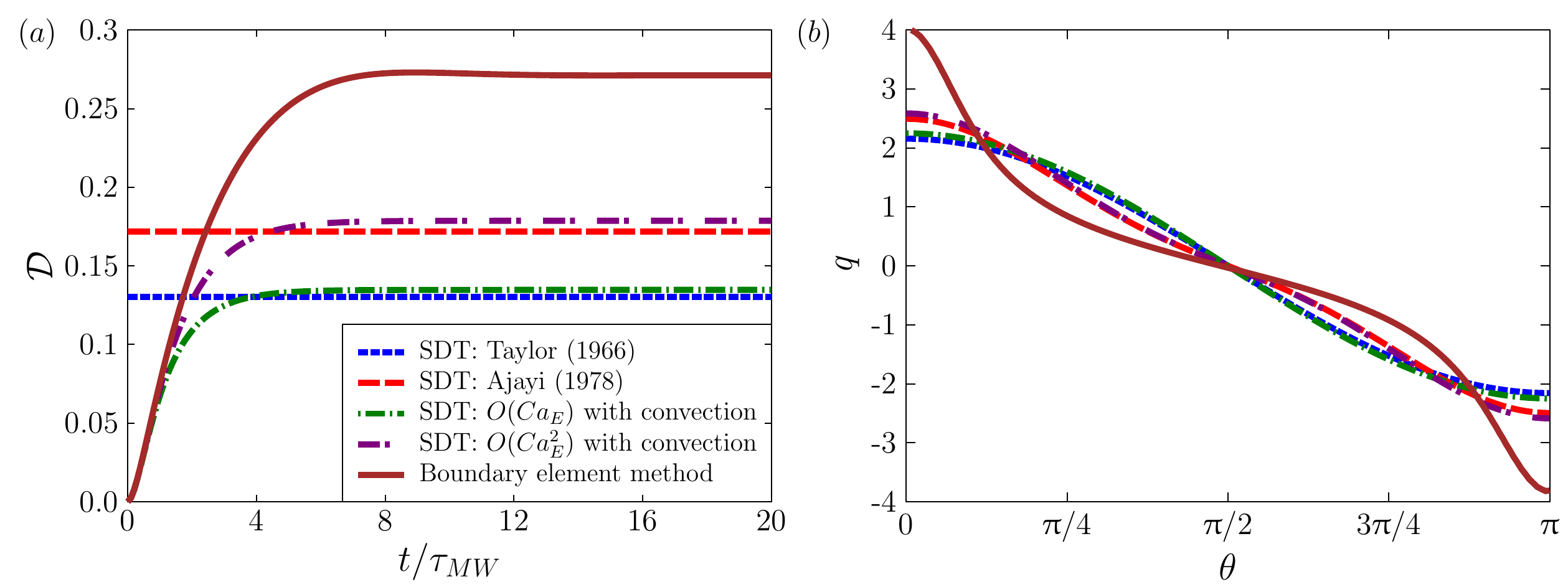}
\caption{(color online)  ($a$) Deformation parameter $\mathcal{D}$ as a function of time for  the parameters of system 3, which correspond to a steady prolate shape. ($b$) Steady interfacial charge profile.}
\label{fig:prolate}
\end{figure}

The case of prolate deformations is illustrated in figure~\ref{fig:prolate} using the parameters of system~3. In this case, the drop deformation increases monotonically with time. The steady deformation parameter obtained by simulations with $Ma=1$ is $\mathcal{D}=0.27$, which slightly exceeds the value of $\mathcal{D}=0.22$ found by  \citet{lac2007}, who neglected charge convection ($Ma\rightarrow \infty$); the experiments of \citet{ha2000a}, for which the value of $Ma$ is unknown, reported a deformation of $\mathcal{D}=0.25$. Our small deformation theory only provides a modest improvement at steady state over the predictions of \cite{taylor1966} and \citet{ajayi1978}, again confirming that nonlinear charge convection has a weaker effect for prolate drops. This again can be rationalized by considering the interfacial charge profile in figure~\ref{fig:prolate}($b$): convection by the flow is seen to cause charge accumulation at the drop poles, and thus does not cause any discontinuity as in the oblate case. Instead, the charge profile remains relatively smooth and therefore can be reasonably well approximated using Legendre polynomials. 

\begin{figure}
\centering
\includegraphics[width=13.1cm]{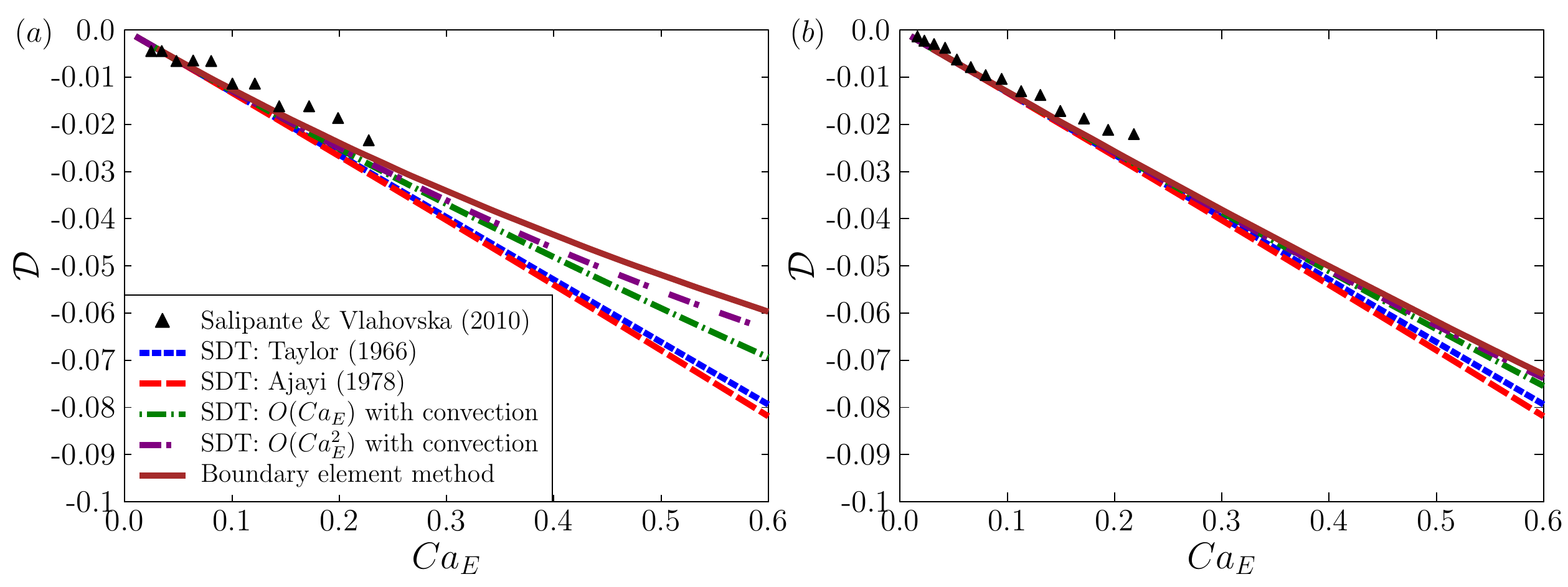}
\caption{(color online) Steady drop deformation $\mathcal{D}$ for the parameters of: ($a$) system 2a, and ($b$) system 2b. The various models are compared to the experimental measurements of \citet{salipante2010}.}
\label{fig:steadydef}
\end{figure}

As a final test, we compare our theoretical and numerical predictions for the steady drop shapes with the experimental results of \citet{salipante2010} for systems 2a and 2b in figure~\ref{fig:steadydef}. The experimental systems used two different drop sizes but identical material properties. At a given value of the electric capillary number $Ca_E$, increasing drop size is equivalent to decreasing the electric field or increasing the Mason number $Ma$, which reduces the effect of charge convection. Charge convection is therefore more significant in figure~\ref{fig:schematic}($a$) for the smaller drop size, and indeed departures of our numerical and theoretical results from the small-deformation theories of \cite{taylor1966} and \citet{ajayi1978} are more significant in this case. In both cases, our model performs quite well at predicting the steady drop shape, but still slightly overpredicts the experimental values especially as $Ca_E$ increases; nonetheless the agreement is noticeably better than previous models.  \vspace{-0.15cm}

\section{Concluding remarks\label{sec:conclusions}}

In summary, we have developed a small-deformation theory for the complete Melcher-Taylor leaky dielectric model including the non-linear charge convection term. A domain perturbation method based on spherical harmonics valid for small deviations from sphericity was employed to represent the drop shape up to second order in electric capillary number ${O}(Ca_E^2)$. The zeroth- and first-order electric and flow fields were solved for using multipole expansions. On making the appropriate assumptions, we were able to recover the previous theoretical models \citep{taylor1966,ajayi1978,esmaeeli2011,lanauze2013}. The discrepancy of Ajayi's second-order theory predicting drop deformation more inaccurately than Taylor's first-order theory in the case of oblate drops was resolved by including charge convection in the theoretical model. Retention of transient charge relaxation and shape deformation was also shown to be critical in order to accurately capture the transient non-monotonic drop deformation, as we validated by comparison with both numerical simulations and existing experimental data.

While our second-order theory showed good agreement with simulations and experiments, departures become significant with increasing electric field strength as deformations become larger. While possible in principle, extending the theory to include higher-order corrections in $Ca_E$ is exceedingly difficult due to the non-linearities in the governing equations. The problem of capturing large deformations in a theoretical model would likely be better addressed using spheroidal coordinates as in the previous work of \citet{zhang2013}, though this method has yet to be adapted to include charge convection. One should also note that the present study is limited to axisymmetric drop deformations. In strong electric fields, experiments have demonstrated the existence of a symmetry-breaking bifurcation leading to Quincke electrorotation \citep{salipante2010,salipante2013,he2013}, which is characterized by non-axisymmetric shapes and a primarily rotational flow. Such effects cannot be captured by the theory and simulations presented herein. From a theoretical standpoint, a fully three-dimensional model would preclude the simple use of a Stokes streamfunction as done in \S \ref{sec:flow} for the solution of the flow problem, which could instead by obtained using Lamb's general solution of the Stokes equations \citep{kim2013}. Such a model would also be useful for the description of pair interactions between widely separated drops using the method of reflections, in a similar manner as in the previous work of \cite{anderson85} for thermocapillary motion of drops, or as in our previous theory for electrohydrodynamic interactions between rigid spheres \citep{das2013}; the understanding of such interactions could then pave the way for  dilute suspension theories for electrohydrodynamics of multiple drops. Lastly, three-dimensional boundary element simulations would also be of great use to describe large deformations and electrorotation in strong fields and are the subject of our current work.  \vspace{-0.15cm}

\section*{Acknowledgements}
The authors thank Petia Vlahovska and Paul Salipante for comments and suggestions, Aditya Khair and Javier Lanauze for useful discussions and for sharing their experimental data, and Lorenzo Rossini for help with generating movies of drop dynamics. Acknowledgment is made to the Donors of the American Chemical Society Petroleum Research Fund for partial support of this research through grant 53240-ND9.

\appendix
\section{Axisymmetric boundary element method}

We outline the numerical method used in \S \ref{sec:results} for the solution of the full nonlinear problem in axisymmetric geometry based on boundary integral equations \citep{jaswon1963,symm1963}. The method shares similarities with that of \cite{lanauze2015} but makes use of a finite-volume algorithm for the solution of the charge convection equation. We first solve Laplace's equation for the electric potential using a single-layer potential \citep{sherwood1988,baygents1998,lac2007, lanauze2015}, yielding the integral equation
\begin{align}
\varphi(\boldsymbol{x}_{0})=-\boldsymbol{x}_0\bcdot \boldsymbol{E}_{0}+\int_{C} \llbracket {E}^{n}(\boldsymbol{x})\rrbracket \, \mathcal{G}^{a}(\boldsymbol{x}_0;\boldsymbol{x})\,\mathrm{d}s(\boldsymbol{x}),\label{eq:singlelayer}
\end{align}
where $C$ is the one-dimensional curve describing the drop shape, which is parametrized by arclength $s$. Equation (\ref{eq:singlelayer}) is valid for any location of the evaluation point $\boldsymbol{x}_0$ on the drop surface $C$ or in either of the fluid domains $V$ and $\bar{V}$. It involves the axisymmetric Green's function for Laplace's equation, which is obtained by integration of the three-dimensional free-space Green's function over the azimuthal direction:
\begin{equation}
\mathcal{G}^{a}(\boldsymbol{x}_{0};\boldsymbol{x})=\int_{0}^{2\uppi}\frac{\mathrm{d}\phi}{4\upi r}, \qquad \mbox{where}\quad r=|\boldsymbol{r}|=|\boldsymbol{x}_0-\boldsymbol{x}|.
\end{equation}
Knowledge of the single-layer potential density $\llbracket {E}^{n}\rrbracket$ therefore allows determination of the electric potential anywhere in space by simple integration, which prompts us to seek an equation for $\llbracket {E}^{n}\rrbracket$ in terms of the charge density $q$. To this end, we first take the gradient of equation (\ref{eq:singlelayer}) with respect to $\boldsymbol{x}_0$ to obtain integral equations for the electric field in both fluid phases:
\begin{subequations}
\begin{align}
\boldsymbol{E}(\boldsymbol{x}_0)=\boldsymbol{E}_{0}-\int_{C} \llbracket {E}^{n}(\boldsymbol{x})\rrbracket \bnabla_{0}\mathcal{G}^{a}\,\mathrm{d}s(\boldsymbol{x}) \quad \mbox{for}\,\,\,\boldsymbol{x}_{0}\in V, \\
\bar{\boldsymbol{E}}(\boldsymbol{x}_0)=\boldsymbol{E}_{0}-\int_{C} \llbracket {E}^{n}(\boldsymbol{x})\rrbracket \bnabla_{0}\mathcal{G}^{a}\,\mathrm{d}s(\boldsymbol{x}) \quad \mbox{for}\,\,\,\boldsymbol{x}_{0}\in \bar{V}.
\end{align}
\end{subequations}
The derivative of the Green's function undergoes a discontinuity across the interface, which needs to be accounted for when the evaluation point is on the boundary \citep{pozrikidis2011}, leading to the following expressions on the drop surface:
\begin{subequations}\label{eq:Eintegral}
\begin{align}
\boldsymbol{E}(\boldsymbol{x}_0)=\boldsymbol{E}_{0}-\int_{C} \llbracket {E}^{n}(\boldsymbol{x})\rrbracket \bnabla_{0}\mathcal{G}^{a}\,\mathrm{d}s (\boldsymbol{x})+\tfrac{1}{2} \llbracket {E}^{n}(\boldsymbol{x})\rrbracket \boldsymbol{n}(\boldsymbol{x}_0) \quad \mbox{for}\,\,\,\boldsymbol{x}_{0}\in C,\\
\bar{\boldsymbol{E}}(\boldsymbol{x}_0)=\boldsymbol{E}_{0}-\int_{C} \llbracket {E}^{n}(\boldsymbol{x})\rrbracket \bnabla_{0}\mathcal{G}^{a}\,\mathrm{d}s (\boldsymbol{x})-\tfrac{1}{2} \llbracket {E}^{n}(\boldsymbol{x})\rrbracket \boldsymbol{n}(\boldsymbol{x}_0) \quad \mbox{for}\,\,\,\boldsymbol{x}_{0}\in C.
\end{align}
\end{subequations}
These equations are singular at $\boldsymbol{x}=\boldsymbol{x}_0$, though the singularity disappears after taking the dot product with the  normal $\boldsymbol{n}(\boldsymbol{x}_0)$. An integral equation for the jump can then be obtained by summing both equations and combining them with Gauss's law (\ref{eq:gauss}), which is written $q=E^n-Q\bar{E}^n$ in dimensionless form. After manipulations, it reads
\begin{align}
\int_{C}\llbracket {E}^{n}(\boldsymbol{x})\rrbracket [\boldsymbol{n}(\boldsymbol{x}_0)\bcdot \bnabla_{0}\mathcal{G}^{a}]\mathrm{d}s (\boldsymbol{x})-\frac{1+Q}{2(1-Q)}\llbracket {E}^{n}(\boldsymbol{x}_0)\rrbracket=E^n_0(\boldsymbol{x}_0)-\frac{q(\boldsymbol{x}_0)}{1-Q}. \label{eq:jumpintegral}
\end{align}
This can be solved for $\llbracket {E}^{n}\rrbracket$, from which $E^n$ and $\bar{E}^n$ are deduced using Gauss's law as
\begin{equation}
E^n=\frac{q-Q\llbracket {E}^{n}\rrbracket}{1-Q}, \qquad \bar{E}^n=\frac{q-\llbracket {E}^{n}\rrbracket}{1-Q}. \label{eq:EnEnbar}
\end{equation}
The tangential component of the electric field can then be obtained by evaluating equation (\ref{eq:Eintegral}), though care must be taken to treat the integral singularity \citep{sellier2006}. Another approach, which we adopt here, consists in evaluating the potential $\varphi$ using equation (\ref{eq:singlelayer}), which is only weakly singular, and then differentiating it numerically along the curve $C$ to obtain $E^t$. 

Once both normal and tangential components of the electric field are known, they can be used to determine the jump in electric tractions $\llbracket \boldsymbol{f}^E\rrbracket$ using equation (\ref{eq:electricf}), from which we infer the jump in hydrodynamic tractions $\llbracket \boldsymbol{f}^H\rrbracket$ using the stress balance (\ref{eq:dynamicBC}). Hydrodynamic tractions then enter the Stokes boundary integral equation for the fluid velocity $\boldsymbol{v}$ \citep{pozrikidis1992}, which for an axisymmetric domain reads
\begin{align} \label{eq:Stokesintegral}
\begin{split}
\boldsymbol{v}(\boldsymbol{x}_0)=&-\dfrac{1}{2\uppi Ma(1+\lambda)} \int_C  \llbracket \boldsymbol{f}^H\rrbracket \bcdot \mathsfbi{G}^{\,a} (\boldsymbol{x};\boldsymbol{x}_0)\, \mathrm{d}s(\boldsymbol{x}) \\
&+ \dfrac{1-\lambda}{4 \uppi(1+\lambda)} \int_C \boldsymbol{v}(\boldsymbol{x})\bcdot \mathsfbi{T}^{a}(\boldsymbol{x};\boldsymbol{x}_0) \bcdot \boldsymbol{n}(\boldsymbol{x})\, \mathrm{d}s(\boldsymbol{x}),
\end{split}
\end{align}
where $\mathsfbi{G}^{\,a}$ and $\mathsfbi{T}^{a}$ are the axisymmetric Green's functions for the Stokeslet and stresslet, respectively:
\begin{align}
\mathsfbi{G}^{\,a}(\boldsymbol{x};\boldsymbol{x}_0) = \int_{0}^{2\uppi} \left(\frac{\mathsfbi{I}}{r} 
+ \frac{\boldsymbol{r}\boldsymbol{r}}{r^3}\right)\,\mathrm{d}\phi, \quad
\mathsfbi{T}^{a}(\boldsymbol{x};\boldsymbol{x}_0) = \int_{0}^{2\uppi} -6\frac{\boldsymbol{r}\boldsymbol{r}\boldsymbol{r}}{r^5} \,\mathrm{d}\phi,
\end{align}
The exact expressions for these functions are very cumbersome but can be found in \citet{pozrikidis1992,pozrikidis2002}.
The integral equation (\ref{eq:Stokesintegral}), which is valid in both fluid domains and on the interface, can be inverted to determine the interfacial velocity, which is then used to update the drop shape and charge distribution. 

The complete algorithm can be summarized as follows:
\begin{enumerate}
\item Given a surface charge distribution $q(\boldsymbol{x})$, compute $\llbracket E^n\rrbracket$, $E^n$, and $\bar{E}^n$ by solution of the integral equation (\ref{eq:jumpintegral}) together with equation (\ref{eq:EnEnbar}).
\item Determine the surface potential $\varphi$ by evaluation of equation (\ref{eq:singlelayer}).
\item Differentiate the surface potential $\varphi$ numerically along the interface to obtain the tangential electric field $\boldsymbol{E}^t=-\bnabla_s \varphi$. 
\item Knowing both components of the electric field, calculate the jump in electric tractions $\llbracket \boldsymbol{f}^E\rrbracket$ using equation (\ref{eq:electricf}), and use it to determine the jump in hydrodynamic tractions $\llbracket \boldsymbol{f}^H\rrbracket$ using the stress balance (\ref{eq:dynamicBC}).
\item Solve the Stokes boundary integral equation (\ref{eq:Stokesintegral}) to obtain the interfacial  velocity.
\item Update the charge distribution $q(\boldsymbol{x})$ by time marching of the charge conservation equation (\ref{eq:chargeeq}) using an explicit scheme.
\item Update the position of the interface by advecting the mesh with the normal component of the interfacial velocity using the same time-marching scheme as in (vi). 
\end{enumerate}

In all simulations, the drop shape is taken to be initially spherical, and the initial surface charge is uniformly zero. We use spline interpolation to represent the shape of the interface, which allows for an easy and accurate determination of geometric properties such as the normal and tangential vectors and surface curvature, and for accurate evaluation of surface integrals. The charge conservation equation, however, is discretized using a finite-volume scheme \citep{leveque2002}, which has better conservation properties and is also more adequate for capturing sharp gradients as arise in strong fields (figures~\ref{fig:1b} and \ref{fig:1c}); this distinguishes our method from that of \citet{lanauze2015}, which uses splines for both the drop shape and surface charge distribution.

\bibliographystyle{jfm}
\bibliography{smalldeftheory_final.bbl}

\begin{thebibliography}{46}
\expandafter\ifx\csname natexlab\endcsname\relax\def\natexlab#1{#1}\fi
\def\au#1{#1} \def\ed#1{#1} \def\yr#1{#1}\def\at#1{#1}\def\jt#1{\textit{#1}}
  \def\bt#1{#1}\def\bvol#1{\textbf{#1}} \def\vol#1{#1} \def\pg#1{#1}
  \def\publ#1{#1}\def\arxiv#1{#1}\def\org#1{#1}\def\st#1{\textit{#1}}

\bibitem[Abramowitz \& Stegun(1972)]{abramowitz72}
{\sc \au{Abramowitz, M.} \& \au{Stegun, I.~A.}} \yr{1972} {\em Handbook of
  Mathematical Functions: with Formulas, Graphs, and Mathematical Tables\/}.
  \publ{Dover}.

\bibitem[Ajayi(1978)]{ajayi1978}
{\sc \au{Ajayi, O.~O.}} \yr{1978}  \at{A note on {T}aylor's electrohydrodynamic
  theory}.  \jt{Proc. R. Soc. Lond. A}  \bvol{364},  \pg{499--507}.

\bibitem[Allan \& Mason(1962)]{allan1962}
{\sc \au{Allan, R.~S.} \& \au{Mason, S.~G.}} \yr{1962}  \at{Particle behaviour
  in shear and electric fields. {I}. {D}eformation and burst of fluid drops}.
  \jt{Proc. R. Soc. Lond. A}  \bvol{267},  \pg{45--61}.

\bibitem[Anderson(1985)]{anderson85}
{\sc \au{Anderson, J.~L.}} \yr{1985}  \at{Droplet interactions in
  thermocapillary motion}.  \jt{Intl. J. Multiphase Flow}  \bvol{11},
  \pg{813--824}.

\bibitem[Bandopadhyay {\em et~al.\/}(2016)Bandopadhyay, Mandal, Kishore \&
  Chakraborty]{bandopadhyay2016}
{\sc \au{Bandopadhyay, A.}, \au{Mandal, S.}, \au{Kishore, N.~K.} \&
  \au{Chakraborty, S.}} \yr{2016}  \at{Uniform electric-field-induced lateral
  migration of a sedimenting drop}.  \jt{J. Fluid Mech.}  \bvol{792},
  \pg{553--589}.

\bibitem[Basaran {\em et~al.\/}(2013)Basaran, Gao \& Bhat]{basaran2013}
{\sc \au{Basaran, O.~A.}, \au{Gao, H.} \& \au{Bhat, P.~P.}} \yr{2013}
  \at{Nonstandard inkjets}.  \jt{Annu. Rev. Fluid Mech.}  \bvol{45},
  \pg{85--113}.

\bibitem[Baygents {\em et~al.\/}(1998)Baygents, Rivette \& Stone]{baygents1998}
{\sc \au{Baygents, J.~C.}, \au{Rivette, N.~J.} \& \au{Stone, H.~A.}} \yr{1998}
  \at{Electrohydrodynamic deformation and interaction of drop pairs}.  \jt{J.
  Fluid Mech.}  \bvol{368},  \pg{359--375}.

\bibitem[Castellanos(2014)]{castellanos2014}
{\sc \au{Castellanos, A.}} \yr{2014} {\em Electrohydrodynamics\/}.
  \publ{Springer}.

\bibitem[Das \& Saintillan(2013)]{das2013}
{\sc \au{Das, D.} \& \au{Saintillan, D.}} \yr{2013}  \at{Electrohydrodynamic
  interaction of spherical particles under {Q}uincke rotation}.  \jt{Phys. Rev.
  E}  \bvol{87},  \pg{043014}.

\bibitem[Esmaeeli \& Sharifi(2011)]{esmaeeli2011}
{\sc \au{Esmaeeli, A.} \& \au{Sharifi, P.}} \yr{2011}  \at{Transient
  electrohydrodynamics of a liquid drop}.  \jt{Phys. Rev. E}  \bvol{84},
  \pg{036308}.

\bibitem[Feng(1999)]{feng1999}
{\sc \au{Feng, J.~Q.}} \yr{1999}  \at{Electrohydrodynamic behaviour of a drop
  subjected to a steady uniform electric field at finite electric {R}eynolds
  number}.  \jt{Proc. R. Soc. Lond. A}  \bvol{455},  \pg{2245--2269}.

\bibitem[Ha \& Yang(2000)]{ha2000a}
{\sc \au{Ha, J.-W.} \& \au{Yang, S.-M.}} \yr{2000}  \at{Deformation and breakup
  of {N}ewtonian and non-{N}ewtonian conducting drops in an electric field}.
  \jt{J. Fluid Mech.}  \bvol{405},  \pg{131--156}.

\bibitem[Harris \& O'Konski(1957)]{harris1957}
{\sc \au{Harris, F.~E.} \& \au{O'Konski, C.~T.}} \yr{1957}  \at{Dielectric
  properties of aqueous ionic solutions at microwave frequencies}.  \jt{J.
  Phys. Chem.}  \bvol{61},  \pg{310--319}.

\bibitem[Haywood {\em et~al.\/}(1991)Haywood, Renksizbulut \&
  Raithby]{haywood1991}
{\sc \au{Haywood, R.~J.}, \au{Renksizbulut, M.} \& \au{Raithby, G.~D.}}
  \yr{1991}  \at{Transient deformation of freely-suspended liquid droplets in
  electrostatic fields}.  \jt{AIChE J.}  \bvol{37},  \pg{1305--1317}.

\bibitem[He {\em et~al.\/}(2013)He, Salipante \& Vlahovska]{he2013}
{\sc \au{He, H.}, \au{Salipante, P.~F.} \& \au{Vlahovska, P.~M.}} \yr{2013}
  \at{Electrorotation of a viscous droplet in a uniform direct current electric
  field}.  \jt{Phys. Fluids}  \bvol{25},  \pg{032106}.

\bibitem[Huang {\em et~al.\/}(2003)Huang, Zhang, Kotaki \&
  Ramakrishna]{huang2003}
{\sc \au{Huang, Z.-M.}, \au{Zhang, Y.-Z.}, \au{Kotaki, M.} \& \au{Ramakrishna,
  S.}} \yr{2003}  \at{A review on polymer nanofibers by electrospinning and
  their applications in nanocomposites}.  \jt{Compos. Sci. Technol.}
  \bvol{63},  \pg{2223--2253}.

\bibitem[Jaswon(1963)]{jaswon1963}
{\sc \au{Jaswon, M.~A.}} \yr{1963}  \at{Integral equation methods in potential
  theory. {I}}.  \jt{Proc. R. Soc. Lond. A}  \bvol{275},  \pg{23--32}.

\bibitem[Kim \& Karrila(2013)]{kim2013}
{\sc \au{Kim, S.} \& \au{Karrila, S.~J.}} \yr{2013} {\em Microhydrodynamics:
  Principles and Selected Applications\/}.  \publ{Dover}.

\bibitem[Lac \& Homsy(2007)]{lac2007}
{\sc \au{Lac, E.} \& \au{Homsy, G.~M.}} \yr{2007}  \at{Axisymmetric deformation
  and stability of a viscous drop in a steady electric field}.  \jt{J. Fluid
  Mech.}  \bvol{590},  \pg{239--264}.

\bibitem[Lanauze {\em et~al.\/}(2013)Lanauze, Walker \& Khair]{lanauze2013}
{\sc \au{Lanauze, J.~A.}, \au{Walker, L.~M.} \& \au{Khair, A.~S.}} \yr{2013}
  \at{The influence of inertia and charge relaxation on electrohydrodynamic
  drop deformation}.  \jt{Phys. Fluids}  \bvol{25},  \pg{112101}.

\bibitem[Lanauze {\em et~al.\/}(2015)Lanauze, Walker \& Khair]{lanauze2015}
{\sc \au{Lanauze, J.~A.}, \au{Walker, L.~M.} \& \au{Khair, A.~S.}} \yr{2015}
  \at{Nonlinear electrohydrodynamics of slightly deformed oblate drops}.
  \jt{J. Fluid Mech.}  \bvol{774},  \pg{245--266}.

\bibitem[Laser \& Santiago(2004)]{laser2004}
{\sc \au{Laser, D.~J.} \& \au{Santiago, J.~G.}} \yr{2004}  \at{A review of
  micropumps}.  \jt{J. Micromech. Microengng.}  \bvol{14},  \pg{R35}.

\bibitem[LeVeque(2002)]{leveque2002}
{\sc \au{LeVeque, R.~J.}} \yr{2002} {\em Finite Volume Methods for Hyperbolic
  Problems\/}.  \publ{Cambridge University Press}.

\bibitem[Melcher \& Taylor(1969)]{melcher1969}
{\sc \au{Melcher, J.~R.} \& \au{Taylor, G.~I.}} \yr{1969}
  \at{Electrohydrodynamics: a review of the role of interfacial shear
  stresses}.  \jt{Annu. Rev. Fluid Mech.}  \bvol{1},  \pg{111--146}.

\bibitem[Moriya {\em et~al.\/}(1986)Moriya, Adachi \& Kotaka]{moriya1986}
{\sc \au{Moriya, S.}, \au{Adachi, K.} \& \au{Kotaka, T.}} \yr{1986}
  \at{Deformation of droplets suspended in viscous media in an electric field.
  {I}. {R}ate of deformation}.  \jt{Langmuir}  \bvol{2},  \pg{155--160}.

\bibitem[O'Konski \& Thacher(1953)]{konski1953}
{\sc \au{O'Konski, C.~T.} \& \au{Thacher, H.~C.}} \yr{1953}  \at{The distortion
  of aerosol droplets by an electric field}.  \jt{J. Phys. Chem.}  \bvol{57},
  \pg{955--958}.

\bibitem[Park {\em et~al.\/}(2007)Park, Hardy, Kang, Barton, Adair,
  Mukhopadhyay, Lee, Strano, Alleyne, Georgiadis, F. \& R.]{park2007}
{\sc \au{Park, J.-U.}, \au{Hardy, M.}, \au{Kang, S.~J.}, \au{Barton, K.},
  \au{Adair, K.}, \au{Mukhopadhyay, D.~K.}, \au{Lee, C.~Y.}, \au{Strano,
  M.~S.}, \au{Alleyne, A.~G.}, \au{Georgiadis, J.~G.}, \au{F., Placid~M.} \&
  \au{R., John~A.}} \yr{2007}  \at{High-resolution electrohydrodynamic jet
  printing}.  \jt{Nat. Mater.}  \bvol{6},  \pg{782--789}.

\bibitem[Pozrikidis(1992)]{pozrikidis1992}
{\sc \au{Pozrikidis, C.}} \yr{1992} {\em Boundary Integral and Singularity
  Methods for Linearized Viscous Flow\/}.  \publ{Cambridge University Press}.

\bibitem[Pozrikidis(2002)]{pozrikidis2002}
{\sc \au{Pozrikidis, C.}} \yr{2002} {\em A Practical Guide to Boundary Element
  Methods with the Software Library BEMLIB\/}.  \publ{CRC Press}.

\bibitem[Pozrikidis(2011)]{pozrikidis2011}
{\sc \au{Pozrikidis, C.}} \yr{2011} {\em Introduction to Theoretical and
  Computational Fluid Dynamics\/}.  \publ{Oxford University Press}.

\bibitem[Rallison(1984)]{rallison1984}
{\sc \au{Rallison, J.M.}} \yr{1984}  \at{The deformation of small viscous drops
  and bubbles in shear flows}.  \jt{Annu. Rev. Fluid Mech.}  \bvol{16},
  \pg{45--66}.

\bibitem[Salipante \& Vlahovska(2010)]{salipante2010}
{\sc \au{Salipante, P.~F.} \& \au{Vlahovska, P.~M.}} \yr{2010}
  \at{Electrohydrodynamics of drops in strong uniform dc electric fields}.
  \jt{Phys. Fluids}  \bvol{22},  \pg{112110}.

\bibitem[Salipante \& Vlahovska(2013)]{salipante2013}
{\sc \au{Salipante, P.~F.} \& \au{Vlahovska, P.~M.}} \yr{2013}
  \at{Electrohydrodynamic rotations of a viscous droplet}.  \jt{Phys. Rev. E}
  \bvol{88},  \pg{043003}.

\bibitem[Saville(1997)]{saville1997}
{\sc \au{Saville, D.~A.}} \yr{1997}  \at{Electrohydrodynamics: the
  {T}aylor-{M}elcher leaky dielectric model}.  \jt{Annu. Rev. Fluid Mech.}
  \bvol{29},  \pg{27--64}.

\bibitem[Scott(1989)]{scott1989}
{\sc \au{Scott, T.~C.}} \yr{1989}  \at{Use of electric fields in solvent
  extraction: a review and prospectus}.  \jt{Sep. Purif. Meth.}  \bvol{18},
  \pg{65--109}.

\bibitem[Sellier(2006)]{sellier2006}
{\sc \au{Sellier, A.}} \yr{2006}  \at{On the computation of the derivatives of
  potentials on a boundary by using boundary-integral equations}.  \jt{Comput.
  Methods Appl. Mech. Eng.}  \bvol{196},  \pg{489--501}.

\bibitem[Sherwood(1988)]{sherwood1988}
{\sc \au{Sherwood, J.~D.}} \yr{1988}  \at{Breakup of fluid droplets in electric
  and magnetic fields}.  \jt{J. Fluid Mech.}  \bvol{188},  \pg{133--146}.

\bibitem[Shkadov \& Shutov(2002)]{shkadov2002}
{\sc \au{Shkadov, V.~Y.} \& \au{Shutov, A.~A.}} \yr{2002}  \at{Drop and bubble
  deformation in an electric field}.  \jt{Fluid Dyn.}  \bvol{37},
  \pg{713--724}.

\bibitem[Shutov(2002)]{shutov2002}
{\sc \au{Shutov, A.~A.}} \yr{2002}  \at{The shape of a drop in a constant
  electric field}.  \jt{Tech. Phys.}  \bvol{47},  \pg{1501--1508}.

\bibitem[Supeene {\em et~al.\/}(2008)Supeene, Koch \&
  Bhattacharjee]{supeene2008}
{\sc \au{Supeene, G.}, \au{Koch, C.~R.} \& \au{Bhattacharjee, S.}} \yr{2008}
  \at{Deformation of a droplet in an electric field: Nonlinear transient
  response in perfect and leaky dielectric media}.  \jt{J. Colloid Interface
  Sci.}  \bvol{318},  \pg{463--476}.

\bibitem[Symm(1963)]{symm1963}
{\sc \au{Symm, G.~T.}} \yr{1963}  \at{Integral equation methods in potential
  theory. {II}}.  \jt{Proc. R. Soc. Lond. A}  \bvol{275},  \pg{33--46}.

\bibitem[Taylor(1964)]{taylor1964}
{\sc \au{Taylor, G.~I.}} \yr{1964}  \at{Disintegration of water drops in an
  electric field}.  \jt{Proc. R. Soc. Lond. A}  \bvol{280},  \pg{383--397}.

\bibitem[Taylor(1966)]{taylor1966}
{\sc \au{Taylor, G.~I.}} \yr{1966}  \at{Studies in electrohydrodynamics. {I}.
  {T}he circulation produced in a drop by electrical field}.  \jt{Proc. R. Soc.
  Lond. A}  \bvol{291},  \pg{159--166}.

\bibitem[Taylor(1969)]{taylor1969}
{\sc \au{Taylor, G.~I.}} \yr{1969}  \at{Electrically driven jets}.  \jt{Proc.
  R. Soc. Lond. A}  \bvol{313},  \pg{453--475}.

\bibitem[Wilson \& Taylor(1925)]{wilson1925}
{\sc \au{Wilson, C. T.~R.} \& \au{Taylor, G.~I.}} \yr{1925}  \at{The bursting
  of soap-bubbles in a uniform electric field}.  \jt{Math. Proc. Cambridge
  Philos. Soc.}  \bvol{22},  \pg{728--730}.

\bibitem[Zhang {\em et~al.\/}(2013)Zhang, Zahn \& Lin]{zhang2013}
{\sc \au{Zhang, J.}, \au{Zahn, J.~D.} \& \au{Lin, H.}} \yr{2013}  \at{Transient
  solution for droplet deformation under electric fields}.  \jt{Phys. Rev. E}
  \bvol{87},  \pg{043008}.

\end{thebibliography}

\end{document}